\documentclass[journal,10pt]{IEEEtran}
\usepackage[nospace,nocompress]{cite}

\ifCLASSINFOpdf
   \usepackage[pdftex]{graphicx}
   \DeclareGraphicsExtensions{.pdf,.jpeg,.png}
\else
   \usepackage[dvips]{graphicx}
   \DeclareGraphicsExtensions{.eps}
\fi
\usepackage[cmex10]{amsmath}
\usepackage{amssymb}
\usepackage{bm}
\usepackage{nccmath}

\usepackage{caption}

\DeclareMathOperator{\Span}{span}
\DeclareMathOperator{\Vol}{vol}
\DeclareMathOperator{\Rank}{rank}
\newtheorem{Def}{\textbf{Definition}}
\newtheorem{Therm}{\textbf{Theorem}}
\newtheorem{Lemma}{\textbf{Lemma}}

\newtheorem{Prop}{\textbf{Proposition}}

\begin{document}
%
\title{A TDOA technique with Super-Resolution based on the Volume Cross-Correlation Function}
%
%
%

\author{Hailong Shi, ~\IEEEmembership{Student Member, IEEE}
        Hao Zhang, ~\IEEEmembership{Member, IEEE}
        and~Xiqin~Wang ~\IEEEmembership{Member, IEEE}
\thanks{H. Shi, H. Zhang, and X. Wang are with the Dept. Electronic Engineering, Tsinghua University.  E-mail: shl06@mails.tsinghua.edu.cn, haozhang@tsinghua.edu.cn, and wangxq\_ee@tsinghua.edu.cn.}
\thanks{
	This work is supported in part by the National Natural Science Foundation of China under Grants 61422110, 41271011, and 61201356, also in part by the Program for New Century Excellent Talents in University under Grant NCET-11-0270, and the Tsinghua University Initiative Scientific Research Program. }
}

\maketitle

\begin{abstract}
Time Difference of Arrival (TDOA) is widely used in wireless localization systems. Among the enormous approaches of TDOA, high resolution TDOA algorithms have drawn much attention for its ability to resolve closely spaced signal delays in multipath environment. However, the state-of-art high resolution TDOA algorithms still have performance weakness on resolving time delays in a wireless channel with dense multipath effect. In this paper, we propose a novel TDOA algorithm with super resolution based on a multi-dimensional cross-correlation function: the Volume Cross-Correlation Function (VCC). Theoretical analyses are provided to justify the feasibility of The proposed TDOA algorithm, and numerical simulations also show an excellent time resolution capability of the algorithm in multipath environment.

\end{abstract}


%
\IEEEpeerreviewmaketitle

\section{Introduction}
\par
Among the tremendous amount of source localization techniques \cite{reed1998overview,caffery1998subscriber,gezici2008survey}, TDOA based techniques are widely used in wireless communication \cite{reed1998overview,cong2002hybrid}, indoor microphone positioning \cite{benesty2008microphone}, wireless sensor network \cite{mao2007wireless}, passive localization system \cite{quazi1981overview,dersan2002passive}, and sonar \cite{huang2001real}.
Since traditional TDOA methods, such as the Generalized Cross-Correlation algorithm (GCC) \cite{knapp1976generalized}, have limited time resolution and can not resolve the TDOA of multipath signals with close delays, many high resolution TDOA algorithms have been proposed recently to deal with the scenario where signals from different paths have close delays. 
\par There are mainly three branches of high resolution TDOA algorithms: one is the optimal maximal likelihood (ML) time delay estimators using techniques like expectation maximization (EM)\cite{feder1988parameter}, or importance sampling\cite{masmoudi2013maximum,masmoudi2011non}; another branch is the super resolution TDOA algorithms based on subspace methods \cite{zi1982new,ge2007super,saarnisaari1997tls}; the third branch is the high resolution TDOA estimation methods using sparse recovery algorithms based on $\ell_1$ optimization\cite{comsa2011source,comsa2011time}. Except for those main branches, some delay estimation techniques that have super resolution and ability of dealing with multipath environment, such as the technique of time delay estimation from low-rate samples over a union of subspaces\cite{gedalyahu2010time,gedalyahu2011multichannel} can also be adapted to TDOA estimation.

As we know, 
improving the time resolution and enhancing the ability of identifying each multipath TDOA are two major tasks concerned in design of TDOA techniques. In this paper, we are going to propose a highly efficient TDOA algorithm, which has strong ability to resolve multipath TDOAs, based on a novel multi-dimensional cross-correlation function, named the Volume Cross-Correlation function (VCC). This VCC function takes two matrices (which represent subspaces), instead of two vectors, as arguments. It calculates the geometrical volume of the high dimensional parallelotope spanned by column vectors of these two matrices. It can be regarded as a generalized distance measure of the two subspaces spanned by columns of each input matrix. In our method, the received signal is formulated as deterministic signal with unknown linear subspace structure contaminated by random noise. Then this unknown subspace is extracted from noise through singular value decomposition of some data matrix, such procedure is actually a denoising process  commonly seen in modern signal processing. Afterwards the VCC function is calculated with inputs being the basis for the estimated subspace. Finally the corresponding TDOA estimation is indicated by the zeros (or equivalently, the peaks of its reciprocal) of the VCC function.

In order to analyze the performance of the proposed TDOA algorithm, we choose the passive localization system as a typical application scenario. In our analysis, the received signals commonly encountered in passive localization systems are divided into two different categories: the slowly changing subspace signal and the fast changing subspace signal. The slowly changing subspace signal means the subspace structure of the signal remains unchanged during the time interval of a large amount of observations. 
As for the fast changing subspace signal, contrast to the term "slowly changing", it refers to the circumstance that the subspace structure are changing among different observations; therefore there is only a single observation available to estimate the current signal subspace. 
The two signal categories will cover most wireless signals encountered in passive localization systems.

The rest of this paper will be as follows. In section II, we give the problem formulation, as well as the definition of VCC function. In section III and section IV, we propose and analyze our proposed TDOA algorithm based on two categories of signals, respectively. The performance of our TDOA method is demonstrated through numerical simulations in section V.

\section{Preliminary Introduction}
%
%
%
%

\subsection{TDOA estimation in multipath environment}

In a typical TDOA-based localization system, due to the complicated environment where buildings and vehicles may lead to significant scattering of wireless signals, there might be dense multipath effect in the wireless channel. The received multipath signals will be:

\begin{small}
	\begin{eqnarray}
x_1(t) &=& \sum_{l_1=1}^{L_1} \alpha_{1,l_1} s(t-\tau_{1,l_1}) + w_1(t), \label{Bs1m}\\
x_2(t) &=& \sum_{l_2=1}^{L_2} \alpha_{2,l_2} s(t-\tau_{2,l_2})+ w_2(t),  \label{Bs2m}
\end{eqnarray}
\end{small}

where $\alpha_{1,l_1}$ and $\alpha_{2,l_2}$ are the propagation gains (also known as the channel coefficients) of the $l_1$th (or $l_2$) path along which the signal transmitting from source to receiver 1 and 2, respectively, $\tau_{1,l_1}$ and $\tau_{2,l_2}$ represents the corresponding path delays, $L_1$ and $L_2$ are the number of channel paths. $w_1(t)$ and $w_2(t)$ are noises.

As we know, the task of TDOA estimation is to determine the difference of time delays of the received signal from different receivers. However, from (\ref{Bs1m}) and (\ref{Bs2m}), it can be seen that in a multipath channel, there are theoretically multiple TDOAs which can be resolved. Denote these TDOAs by
\begin{equation}
	\Delta \tau_{l_1,l_2} := \tau_{2,l_2}-\tau_{1,l_1},\ l_1=1,\cdots, L_1, l_2 = 1,\cdots,L_2,\label{MultiTDOA}
\end{equation}
we call these multiple TDOAs as \textit{multipath TDOA}. Although in source localization systems, the direct path TDOA is the only concerned, which is $\Delta \tau_{1,1}= \tau_{2,1}-\tau_{1,1}$, precise estimation of the direct path TDOA $\Delta \tau_{1,1}$ actually requires resolution of every multipath TDOA in (\ref{MultiTDOA}). In other words, because the channel path delays and propagation gains are basically unknown at the receivers, we cannot tell the difference between direct path TDOA and other indirect path TDOAs merely from the received signals. Therefore, we need to resolve every mutipath TDOA, before we pick the direct path TDOA and continue the localization process. 
From this point of view, the primary goal of TDOA localization in multipath environment is to precisely resolve every multipath TDOA shown in (\ref{MultiTDOA}). 
\subsection{The Volume Cross-Correlation Function}
The basic relationship between linear subspaces are generally described by principal angles \cite{miao1992principal}. The principal angle is defined as:
\begin{Def}
	Consider linear subspaces $\mathcal{X}_1$ and $\mathcal{X}_2$, with dimensions $\dim(\mathcal{X}_1)=d_1,\dim(\mathcal{X}_2)=d_2$, denote $m = \min(d_1,d_2)$. The principal angles between subspaces $\mathcal{X}_1$ and $\mathcal{X}_2$, denoted by $0 \leq \theta_1 \leq \cdots \leq \theta_m \leq \pi/2$, are defined recursively as
	\begin{eqnarray}
	 \cos \theta_i = \max_{\bm u_i \in \mathcal{X}_1,\bm v_i \in \mathcal{X}_2}\bm u_i^T \bm v_i, \qquad\nonumber \\
	 \text{subject to} \quad 
	\|\bm u_i\|_2=\|\bm v_i\|_2=1, \\
	\qquad \qquad \qquad \qquad \bm u_i^T \bm u_j = 0, \bm v_i^T \bm v_j=0,
	\nonumber
	\end{eqnarray}
	where $ i = 1,\cdots m, j = 1,\cdots,i-1$.
\end{Def}

The principal angle is an important mathematical tool to depict the relationship between subspaces. Except for playing a key role in deriving the geodesic distance \cite{absil2004riemannian} for Grassmann manifold \cite{absil2004riemannian}\cite{qiu2005unitarily}, principal angle are also used to define various distance metrics of linear subspaces \cite{qiu2005unitarily}. The proposed VCC function in this paper is related with the principal angle.

The Volume Cross-Correlation (VCC) function of two given matrices $\bm X_1 \in \mathbb{C}^{n \times d_1}$ and $\bm X_2 \in \mathbb{C}^{n \times d_2}$ is defined as
\begin{equation}\label{VolCC}
\text{vcc}(\bm X_1, \bm X_2):=\frac{\Vol_{d_1+d_2}([\bm X_1, \bm X_2])}{\Vol_{d_1}(\bm X_1)\Vol_{d_2}(\bm X_2)},
\end{equation}
where $[\bm X_1, \bm X_2]$ means putting the columns of matrices $\bm X_1$ and $\bm X_2$ together into one matrix, and $\Vol_{d}(\bm X)$ denotes
the geometrical volume of matrix $\bm X \in \mathbb{C}^{n\times d}$ with dimension $d$ ($d< n$). It is defined as \cite{ben1992volume}:
\begin{equation}\label{VolumeDef1}
\Vol_d (\bm X):= \prod_{i=1}^d \sigma_i,
\end{equation}
where $\sigma_1 \geq \sigma_2 \geq \cdots \geq \sigma_d \geq 0$ are singular values of matrix $\bm X$. Indeed, $\Vol_d (\bm X)$ is the geometrical volume of $d$ dimensional parallelotope spanned by the column vectors of matrix $\bm X$. 

The relation between volume and principal angles is described by the next proposition from \cite{miao1992principal}:
\begin{Prop}\label{Lemma1}
	Consider two linear subspaces $\mathcal{X}_1$ and $\mathcal{X}_2$ in $\mathbb{R}^N$, their dimensions are $\dim(\mathcal{X}_1)=d_1,\dim(\mathcal{X}_2)=d_2$, and their basis matrices are $\bm X_1 \in \mathbb{R}^{N \times d_1}$ and $\bm X_2 \in \mathbb{R}^{N \times d_2}$, then we have
\begin{equation}\label{VolAng}
\frac{\Vol_{d_1+d_2}([\bm X_1, \bm X_2])}{\Vol_{d_1}(\bm X_1)\Vol_{d_2}(\bm X_2)}= \prod_{j=1}^{\min(d_1,d_2)} \sin \theta_j(\mathcal{X}_1, \mathcal{X}_2),
\end{equation}
where $0 \leq \theta_j (\mathcal X_1, \mathcal X_2)\leq 2\pi, 1\leq j \leq \min(d_1,d_2)$ are the principal angles between $\mathcal{X}_1$ and $\mathcal{X}_2$.
\end{Prop}

(\ref{VolAng}) just indicates that the VCC function (\ref{VolCC})
is actually the product of sines of the principal angles between subspaces $\mathcal{X}_1$ and $\mathcal{X}_2$. Therefore (\ref{VolCC}) can be regarded as a kind of distance measure of subspaces $\mathcal{X}_1$ and $\mathcal{X}_2$. Intuitively, if the subspace $\mathcal{X}_1$ and $\mathcal{X}_2$ are linearly dependent, then $\dim(\mathcal{X}_1 \bigcap \mathcal{X}_2) > 0$. According to the definition of principal angles, there must be a vanishing principal angle $\theta_j (\mathcal X_1, \mathcal X_2)$. That is
$\Vol_{d_1+d_2}([\bm X_1, \bm X_2])/\Vol_{d_1}(\bm X_1)\Vol_{d_2}(\bm X_2)=0$.
On the other hand, if $\mathcal{X}_1$ is perpendicular to $\mathcal{X}_2$, then
$\Vol_{d_1+d_2}([\bm X_1, \bm X_2])/\Vol_{d_1}(\bm X_1)\Vol_{d_2}(\bm X_2)=1$
holds obviously. As a matter of fact, VCC function measures the extent of linear dependency between subspaces \cite{robinson1998separation}, and will be used to derive our TDOA algorithm.

\section{Estimating the TDOA of slowly changing subspace signals using VCC function}
\subsection{The slowly-changing subspace signal}
\par In the passive localization system, information about the wireless channel as well as the source signals are generally unknown by the receivers. Hence TDOA technique is quite suitable for this kind of localization system \cite{ho2008passive,ho2007source}. Firstly, we focus on the category of signals that have a slowly changing subspace structure.
\par
A typical type of signals we encounter in passive localization systems are radar signals radiated by non-cooperative radar transmitters. The common pulse radar waveform can be expressed as the following expression:
\begin{equation}
	s(t) = \sum_{m=-\infty}^{+\infty}\sqrt{P_s}g(t-m T_p),\label{RadarTransmt}
\end{equation}
where $P_s>0$ is the transmitting power of the radar, and $g(t)\in \mathbb{C}$ is the general form of the radar pulse waveform, $T_p$ is the pulse repetition interval (PRI).

We consider the received signal in a sigle PRI, then
in the multipath environment, the received signal from the $i$'th receiver ($i=1,2,\cdots$) would be
\begin{equation}\label{receive_radar}
	x_i(t) = \sum_{l_i=1}^{L_i} \alpha_{i,l_i} \sqrt{P_s}g(t-\tau_{i,l_i}) + w_i(t),\quad t \in (0,T_p),
\end{equation}
where $\alpha_{i,l_i}$ and $\tau_{i,l_i}$ are multipath channel coefficients and path delays, $w_i(t)$ is the Gaussian white noise of $i$'th receiver.
After sampling the received signal with the rate $1/T_s$, then (\ref{receive_radar}) can be written into
\begin{equation}
	y_i(kT_s) =
	\sum_{l_i=1}^{L_i} \alpha_{i,l_i} \sqrt{P_s}g(k T_s-\tau_{i,l_i}) + w_i(kT_s).
	\label{digit_rec_subs}
\end{equation}
Rewrite (\ref{digit_rec_subs}) into a vector form, we have
\begin{equation}
	\bm y_i = \sqrt{P_s} \bm G_i \bm \alpha_i + \bm w_i,\qquad i = 1,2,\cdots,\label{digit_rec_vector}
\end{equation}
where $\bm y_i:= [y_i(0),y_i(T_s),\cdots y_i((N-1) T_s)]\in \mathbb{C}^N$, and
\begin{equation}
	\bm G_i = \left[\bm g_{i,1}, \bm g_{i,2},\cdots,\bm g_{i,L_i}\right] \in \mathbb{C}^{N \times L_i},
\end{equation}
where $\bm g_{i,l} := [g(0\cdot T_s-\tau_{i,l_i}), g(1\cdot T_s-\tau_{i,l_i}),\cdots, g((N-1)\cdot T_s-\tau_{i,l_i})]^T,l=1,\cdots,L_i$.
The vector $\bm \alpha_{i} = [\alpha_{i,1},\cdots \alpha_{i,L_i}]\in \mathbb{C}^{L_i}$ is the channel coefficient vector composed of the maltipath channel gains, and $\bm w_i$ is the noise vector. As is shown in (\ref{digit_rec_vector}), the received radar signal in a multipath channel generally has a deterministic subspace structure with the corresponding subspace $\Span(\bm G)$, i.e., spanned by different time-shift versions of radar waveform $g(t)$.

Except for radar signals, the common linearly modulated wireless communication signals such as DS-CDMA, OFDM, QAM, and others that carry symbols on some periodic pulse shapes, can also be modeled as the signal with a subspace structure in (\ref{digit_rec_vector}) \cite{therrien2000time,simeone2004pilot,fock2002low}. The subspace signal structure is mainly related with the channel's path delays $\tau_{i,l_i}$. As a matter of fact, channel delays are caused by different distances between receivers and signal sources (or reflective objects), and generally signal sources and reflective objects don't have extremely high velocities, therefore channel delays can be generally seen to be constant in a short time. On the other hand, the wireless channel's path gains $\alpha_{i,l_i}$ fluctuates with time, which is caused by channel fading effect. This fact also means that for a time interval long enough for the receiver to obtain relatively large samples of the received signal (according to chapter 2 of \cite{tse2005fundamentals}, the time scale of this interval can be up to 20s in a typical channel scenario), these sample data can be formulated as
\begin{equation}\label{digit_rec_vector_multi}
	\bm y_i^{(j)} =  \bm G_i \bm \alpha_i^{(j)} + \bm w_i^{(j)},\qquad j = 1,2,\cdots,
\end{equation}
where $j$ indicates different observations at different time, or snapshots. Although the channel coefficient vector $\bm \alpha^{(j)}$ is fluctuating with different $j$, the subspace structure determined by matrix $\bm G_i$, will remain almost unchanged. We call this category of signals \textit{the slowly changing subspace signal}, meaning that the subspace structure in (\ref{digit_rec_vector}) changes slowly with time, and can be treated as invariant during the observation interval.
\par The subspace structure $\Span(\bm G_i)$ is unknown to the receivers, but can be estimated from multiple observation data like (\ref{digit_rec_vector_multi}).
\begin{figure}[htbp]
	\centering
	\includegraphics[width=0.45\textwidth]{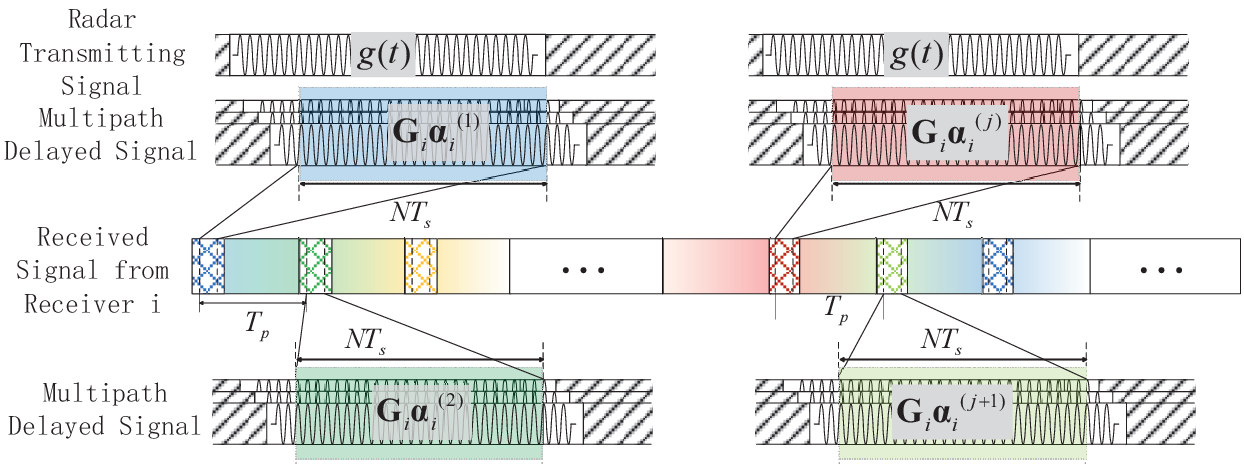}
	\caption{Getting multiple observations of radar signals}
	\label{figurerecv}
\end{figure}
Because a typical radar transmits a pulse waveform repeatedly with a PRI of $T_p$, we can receive multiple snapshots of these multiple radar pulses as in (\ref{digit_rec_vector_multi}) according to $T_p$, which can be estimated by various PRI identification techniques \cite{ata2007deinterleaving,liu2011mutual}. Afterwards the corresponding signal subspace is estimated using the well-known subspace methods like MUSIC, ESPRIT, etc. The process of obtaining multiple observations and estimating subspace $\Span(\bm G_i)$ can be demonstrated in figure \ref{figurerecv}, the gradient change of the background color in figure \ref{figurerecv} represents the fluctuation of channel coefficients $\alpha_{i,l_i}$, they are reasonably assumed to take independent values among different pulses' durations.
\par Denote these multiple data of the received radar signal by
\begin{equation}
\bm y_i^{(1)},\cdots,\bm y_i^{(m)},
\end{equation}
they are then used to evaluate the sampled covariance matrix 
\begin{equation}
\bm{\hat{R}}_i = \frac{1}{m}\sum_{j=1}^{m} \bm y_i^{(j)} \bm (\bm y_i^{(j)})^H,\label{CovEst}
\end{equation}
According to the well-known subspace methods, the signal subspace can be estimated through eigen-decomposition of $\bm{\hat R}_i$. We can firstly evaluate the eigenvalues and eigenvectors of $\bm{\hat R}_i$, then we estimate the dimension of the signal subspace (if the matrix $\bm G_i$ is full rank, the dimension will be $L_i$\footnote{In following analysis, we will assume this full-rankness to be satisfied. Indeed, a sufficient condition will be given to ensure this assumption.}) by analyzing the distribution of eigenvalues; and finally we can separate the eigenvectors of $\bm{\hat R}_i$ into bases for signal subspace as well as bases for noise subspace, The bases for signal subspace are eigenvectors with respect to the $L_i$ largest eigenvalues. As a result, we can write  $\bm{\hat R}_i$ into
\begin{equation}
\bm{\hat R}_i = \bm{U}_{i,s}\bm{\Lambda}_{i,s}\bm{U}_{i,s}^H + \bm{U}_{i,n}\bm{\Lambda}_{i,n}\bm{U}_{i,n}^H,
\end{equation}
where the matrix $\bm{U}_{i,s}$ is the estimated basis matrix for the signal subspace $\Span(\bm G_i)$.
It has been proved that $\Span(\bm{U}_{i,s})$ approximates the signal subspace $\Span(\bm G_i)$ asymptotically for a sufficiently large $m$ \cite{stoica1989music,jeffries1985asymptotic}. Then we will use the estimated basis of signal subspace, i.e., $\bm{U}_{i,s}$, to estimate TDOA using VCC function. 

\subsection{TDOA estimation using VCC function for slowly changing subspace signals: Algorithm}
\par
\hrule
\hspace{1pt}\\
\textbf{Algorithm 1.}
\par \textbf{For $\Delta \tau \in \left((-N+1)T_s,(N-1)T_s\right)$, }
\begin{quote}

\begin{enumerate}
	\setlength\itemsep{0.4em}
	\item Obtain delayed multiple observation data (for $j = 1,\cdots,m$)
	\begin{small}
		 \begin{align}
		 \bm y_1^{(j),[\Delta \tau]} = \hspace{5.2cm}\nonumber \\
		 \left[y_1^{(j)}(0\cdot T_s-\Delta \tau),\cdots,y_1^{(j)}((N-1)T_s-\Delta \tau)\right]^T\nonumber
		 \end{align}
	\end{small}
	and non-delayed observation data
		\begin{small}
			\begin{equation}
		\bm y_2^{(j)} = \left[y_2^{(j)}(0 \cdot T_s),\cdots,y_2^{(j)}((N-1)T_s)\right]^T.\nonumber
		\end{equation}
		\end{small}
	\item Calculate the sampled covariance matrices $\small\bm{\hat R}_1^{[\Delta \tau]}$ and $\small\bm{\hat R}_2$ according to (\ref{CovEst}),
	\item Perform eigenvalue decomposition of the sampled covariance matrix $\small\bm{\hat R}_1^{[\Delta \tau]}$ (and $\small\bm{\hat R}_2$);
	\item Estimate the dimensions of signal subspaces $L_1$ (and $L_2$) based knowledge of the eigenvalues; 
	\item Extract the basis for the signal subspaces $\small\bm{U}_{1,s}^{[\Delta \tau]} \in \mathbb{C}^{N \times L_1}$ (and $\small\bm{U}_{2,s} \in \mathbb{C}^{N \times L_2}$) from the eigenvectors corresponding to the $L_1$ (and $L_2$) largest eigenvalues of $\small\bm{\hat R}_1^{[\Delta \tau]}$ (and $\small\bm{\hat R}_2$):
	\item
	Calculate the reciprocal of VCC function:
\begin{small}
	\begin{equation}
r_{vol}(\Delta \tau):= 1/\Vol_{L_1+L_2}([\bm{U}_{1,s}^{[\Delta \tau]}, \bm{U}_{2,s}]).\nonumber
\end{equation}
\end{small}
\end{enumerate}
\end{quote}
\textbf{Find the peaks of $r_{vol}(\Delta \tau)$ and the corresponding $\Delta \tau$.}
\vspace{2pt}
\hrule
\vspace{4pt}\par
Below are some supplementary remarks: \par
	Remark 1. In our algorithm, the dimensions of signal subspaces are actually the number of channel propagation paths, i.e., $L_i$ in (\ref{receive_radar}), which are generally unknown at the receiver. Therefore the dimensions have to be estimated first. Luckily there are various methods to estimate the dimension of signal subspace from a sampled covariance matrix, such as Akaike Information Criterion (AIC) \cite{akaike1974new} Minimum Description Length (MDL)\cite{wax1985detection}, Bayesian Information Criterion (BIC)\cite{schwarz1978estimating}, Predictive Description Length (PDL)\cite{valaee2004information} and so on. In this paper we just assume the number of channel paths $L_i$ is precisely estimated.
	
	Remark 2.
	The basis matrices $\bm{U}_{1,s}^{[\Delta \tau]}$ and $\bm{U}_{2,s}$ are the estimated bases for the delayed signal subspace $\Span(\bm G_1^{[\Delta \tau]})$ and non-delayed subspace $\Span(\bm G_2)$, respectively. From the previous analysis, we know that
	\begin{equation}\label{subbasis1}
	\bm G_1^{[\Delta \tau]} = \left[\bm g_{1,1}^{[\Delta \tau]},\cdots,\bm g_{1,L_1}^{[\Delta \tau]}\right] \in \mathbb{C}^{N \times L_1},
	\end{equation}
	where $\bm g_{1,l_1}^{[\Delta \tau]}:= [g(0\cdot T_s-\Delta \tau-\tau_{1,l_1}),g(1\cdot T_s-\Delta \tau-\tau_{1,l_1}),$ $\cdots,g((N-1)T_s-\Delta \tau-\tau_{1,l_1})]^T$, and
	\begin{equation}\label{subbasis2}
	\bm G_2 = 
	\left[\bm g_{2,1},\cdots,\bm g_{2,L_2}\right] \in \mathbb{C}^{N \times L_2},
	\end{equation}
	where $\bm g_{2,l_2}:= [g(0\cdot T_s-\tau_{2,l_2}),g(1\cdot T_s-\tau_{2,l_2}),$ $\cdots,g((N-1)T_s-\tau_{2,l_2})]^T$.	
	So the VCC function $\Vol_{L_1+L_2}([\bm{U}_{1,s}^{[\Delta \tau]}, \bm{U}_{2,s}])$ is approximately measuring the linear dependence of subspaces $\Span(\bm G_1^{[\Delta \tau]})$ and $\Span(\bm G_2)$. It is obvious that when $\Delta \tau = \tau_{2,l_2}-\tau_{1,l_1}$, for any pair of $l_1$ and $l_2$, $\Span(\bm G_1^{[\Delta \tau]})$ and $\Span(\bm G_2)$ are linearly dependent, which means $ 1/\Vol_{L_1+L_2}([\bm{U}_{1,s}^{[\Delta \tau]}, \bm{U}_{2,s}])$ will tend to infinity; while on the other hand, when $\Delta \tau \neq \tau_{2,l_2}-\tau_{1,l_1}$
	$\Span(\bm G_1^{[\Delta \tau]})$ and $\Span(\bm G_2)$ may be linearly independent, meaning that $ 1/\Vol_{L_1+L_2}([\bm{U}_{1,s}^{[\Delta \tau]}, \bm{U}_{2,s}])$ would have a finite value. From this observation we could expect that $r_{vol}(\Delta \tau)$ would reach its peak at every $\Delta \tau = \tau_{2,l_2}-\tau_{1,l_1}$. We will give a theoretical guarantee in the next section to validate this observation.
	
	Remark 3. It should be noted that the reciprocal of VCC function $r_{vol}(\Delta \tau)$ is actually a continuous function of variable $\Delta \tau$. So the final step of the proposed algorithm actually involves a continuous one-dimensional parameter search. In practice, when we are manually delaying the signal $\bm y_1^{(j)}$, $\Delta \tau$ can only be integer multiple of $T_s$, i.e., $\Delta \tau$ can only take discrete values related with $T_s$. However fortunately, as long as the signal is sampled at Nyquist rate, we can get the signal $\bm y_1^{(j),[\Delta \tau]}$ for any $\Delta \tau$ without loss by interpolating the original signal until $\Delta \tau$ is on the sampling grid. This interpolation can be a pre-process of the algorithm and is verified by simulation.
	
	Remark 4. Theoretically, we are expecting to resolve every multipath TDOA, i.e.,
	\begin{equation}
	\Delta \tau_{l_2,l_1} := \tau_{2,l_2}-\tau_{1,l_1}, \nonumber
	\end{equation}
	for all $l_1=1,\cdots, L_1,\text{and } l_2 = 1,\cdots,L_2$. It is another independent problem about how can find the TDOA corresponding to the direct channel path, i.e., identifying $\Delta \tau_{1,1} = \tau_{2,1}-\tau_{1,1}$ among all of the multipath TDOA, which is known as the disambiguation of TDOA \cite{scheuing2006disambiguation,scheuing2008disambiguation}. This topic won't be discussed in this paper and will be left for a future work. 	
\subsection{A Theoretical Guarantee for Algorithm 1}
\par Denote the auto-correlation function of radar waveform $g(k\cdot T_s)$ by
\begin{small}
	\begin{equation}
R_g(\tau):=\sum_{k=0}^{N-1} g(k T_s) \cdot g^*(k T_s-\tau), \nonumber
\end{equation}
\end{small}
The auto-correlation function is well known as the ambiguity function of a radar waveform along the zero-Doppler axis, it is an important characteristic of the radar pulse waveform. 
We will show here that, the performance of the proposed algorithm is theoretically guaranteed, and related with the auto-correlation function of the transmitted radar waveform.
A lemma is need firstly:
\begin{Lemma}\label{volRes1_main}
	Consider the matrices $\bm G_1^{[\Delta \tau]}$ and $\bm G_2$ in (\ref{subbasis1}) and (\ref{subbasis2}), 
	if 
	\begin{eqnarray}
		\frac{|R_g(\tau)|}{|R_g(0)|} < \frac{1}{L_1+L_2-1} \text{ for } |\tau| > \Delta \tau^*,\label{ACFconst1}
	\end{eqnarray}
	then
	$\Rank(\bm G_1^{[\Delta \tau]})=L_1,\Rank(\bm G_2)=L_2$.
Here
\begin{equation}
\Delta \tau^* := \min\{\Delta \tau_1,\Delta \tau_2
\},
\end{equation}
and
\begin{small}
	\begin{eqnarray}
\Delta \tau_1 &:=& \min_{1\leq l_1\neq {l_1}' \leq L_1} |\tau_{1,l_1}-\tau_{1,{l_1}'}|,\nonumber \\
\Delta \tau_2 &:=& \min_{1\leq l_2\neq {l_2}' \leq L_2} |\tau_{2,l_2}-\tau_{2,{l_2}'}|.\nonumber
\end{eqnarray}
\end{small}	
\end{Lemma}
\par Lemma \ref{volRes1_main} provides a sufficient condition for the full-rankness of matrices $\bm G_1^{[\Delta \tau]}$ and $\bm G_2$.
Under the condition (\ref{ACFconst1}) of Lemma \ref{volRes1_main}, all the multipath TDOAs $\tau_{2,l_2}-\tau_{1,l_1}$ will theoretically be resolvable by our algorithm, the following theorems will describe the theoretical behavior of your algorithm when Lemma \ref{volRes1_main} holds:

\begin{Therm}\label{volRes1_main3}
		When Lemma \ref{volRes1_main} holds, consider the matrices $\bm G_1^{[\Delta \tau]}$ and $\bm G_2$ in (\ref{subbasis1}) and (\ref{subbasis2}), when for any $1\leq l_1 \leq L_1$ and $1\leq l_2\leq L_2$, $\Delta \tau \neq \tau_{2,l_2} - \tau_{1,l_1}$, given any parameter $\mu$ satisfying
		\begin{equation}
		\mu < \frac{1}{L_1+L_2-1},
		\end{equation}
		if
		\begin{eqnarray}
		\frac{|R_g(\tau)|}{|R_g(0)|} \leq \mu, \text{ for } |\tau| \geq \Delta \tau_{\min},\label{ACFconst2}
		\end{eqnarray}
		then we have
		\begin{equation}\label{volapprox}
		\Vol_{L_1+L_2}([\bm U_{1}^{[\Delta \tau]}, \bm U_{2}])\geq (1-\varepsilon)^{L/2},
		\end{equation}
	 where
	 \begin{equation} 
	 \Delta \tau_{\min}:=\min_{1\leq l_1 \leq L_1,1\leq l_2\leq L_2}|\Delta\tau - \tau_{2,l_2} + \tau_{1,l_1}| ,  
	 \end{equation}
	$L=\min\{L_1,L_2\}$ and the parameter $\varepsilon$ is
	\begin{equation}
	\varepsilon = \frac{L_1L_2\cdot \mu^2}{[1-(L_1-1)\mu][1-(L_2-1)\mu]}.
	\end{equation} 	
\end{Therm}
\begin{Therm}\label{volRes1_main2}
	If there exist $l_1^*$ and $l_2^*$, such that $ \Delta \tau = \tau_{2,l_2^*} - \tau_{1,l_1^*}$, then
	\begin{equation}
	\Vol_{L_1+L_2}([\bm U_{1}^{[\Delta \tau]}, \bm U_{2}])=0. 
	\end{equation}
	The matrices $\bm U_{1}^{[\Delta \tau]}$ and $\bm U_{2}$ here are orthogonal basis matrices for subspaces $\Span(\bm G_1^{[\Delta \tau]})$ and $\Span(\bm G_2)$.
\end{Therm}	

\par Theorems \ref{volRes1_main3} and \ref{volRes1_main2} provide sufficient conditions for the feasibility of our proposed TDOA algorithm. Theorem \ref{volRes1_main2} shows that, when there exist $ l_1^*$ and $l_2^*$, such that 
\begin{equation}
	\Delta \tau = \tau_{2,l_2^*} - \tau_{1,l_1^*},\nonumber
\end{equation}
the VCC function will certainly be zero. This means $1/\Vol_{L_1+L_2}([\bm U_{1}^{[\Delta \tau]}, \bm U_{2}])$ reaches infinity when $\Delta \tau$ equals the TDOA corresponding to path delays $\tau_{2,l_2^*}$ and $\tau_{1,l_1^*}$. On the other hand, according to theorem \ref{volRes1_main3}, when 
\begin{equation}
	 \Delta \tau \neq \tau_{2,l_2} - \tau_{1,l_1}, \nonumber
\end{equation}
as long as the condition (\ref{ACFconst2}) holds, the volume function $\Vol_{L_1+L_2}([\bm U_{1}^{[\Delta \tau]}, \bm U_{2}])$ will be non-zero (because $\varepsilon$ is surely less than $1$ for $0 < \mu < 1/(L_1+L_2-1)$) as shown in (\ref{volapprox}). Because Theorems \ref{volRes1_main3} and  \ref{volRes1_main2} ensure the different values of $\Vol_{L_1+L_2}([\bm U_{1}^{[\Delta \tau]}, \bm U_{2}])$ when $\Delta \tau$ is under different conditions, thus guarantee the feasibility of our proposed algorithm for identifying multipath TDOA.

It is worth noticing that we are using the reciprocal of VCC function, i.e.,
$1/\Vol_{L_1+L_2}([\bm U_{1}^{[\Delta \tau]}, \bm U_{2}])$ to identify the multipath TDOA, because it approximately reaches infinity when $\Delta \tau$ equals one TDOA, and otherwise, remains a finite value. This procedure is similar to the reciprocal technique commonly known in MUSIC method, thus we could conjecture a similar super-resolution characteristic when identifying TDOA. Numerical simulations in section V also show that the VCC function does reveal sharp peaks at the corresponding TDOA locations, thus has super-resolution.

\par If we take a close look at (\ref{ACFconst1}) and (\ref{ACFconst2}), it is obvious that the behavior of our algorithm is related with the autocorrelation function of the radar waveform $g(kT_s)$, i.e., $R_g(\tau)$.  Basically, the capability of identifying different TDOAs is determined by the autocorrelation function $R_g(\tau)$. 
Typically $|R_g(\tau)|$ reaches its maxima at $\tau=0$; and $|R_g(\tau)|$ will eventually (or oscillatorily) decrease as $|\tau|$ increases. So if the radar waveform's autocorrelation function has just one sharp and narrow peak at $\tau=0$, the proposed algorithm will be able to identify different TDOAs that are close to each other; while on the other hand, when $|R_g(\tau)|$ drops slowly as $|\tau|$ increase, 
then for example, 
if there are $l_1^*$ and ${l_1^*}'$ such that $\tau_{1,l_1^*} - \tau_{1,{l_1^*}'}$ are too small to satisfy (\ref{ACFconst1}), TDOAs related with $\tau_{1,l_1^*}$ and $\tau_{1,{l_1^*}'}$ might no longer be identifiable.


As we have discussed, signals with different autocorrelation functions will cause different TDOA estimation performance of our algorithm. Because the width and sharpness of mainlobe of auto-correlation function is actually related with the corresponding signal's bandwidth, wideband radar signals would bring better precision in our TDOA estimation algorithm, simulation will be carried out to support this conclusion. From this point of view, Theorems \ref{volRes1_main2} and \ref{volRes1_main3} coincides with traditional theoretical analyses of TDOA methods like GCC, where the bandwidth of signals is always a key factor influencing the TDOA esimation precision.
	

\section{Estimating the TDOA of fast changing subspace signals using VCC function}

\subsection{The fast changing subspace signal}
 
Contrast to the slowly changing subspace signal model, there are also a large category of signals that don't have a steady subspace structure as in (\ref{digit_rec_vector}). For example, in passive localization systems, FM radio transmitters, TV broadcast stations are usually the signal sources to localize, or are used as the illuminators-of-opportunity to localize a reflective target. Because this category of signals are randomly varying with time and have no repeating waveforms, we cannot get multiple observations of the received signal as in (\ref{digit_rec_vector_multi}) that have the same subspace structure. This category of signals is called the \textit{fast changing subspace signal}. In this case, the received baseband signal from the $i$th receiver is in the form of:
\begin{equation}
x_i(t) = \sum_{l_i=1}^{L_i} \alpha_{i,l_i} s(t-\tau_{i,l_i}) + w_i(t),\ i=1,2,\cdots,\label{fast_sub_sig}
\end{equation}
where $\alpha_{i,l_i}$ and $\tau_{i,l_i}$ are channel's path gain and path delay, respectively. The original transmitted signal $s(t)$ can be FM, PSK or AM signals, etc. 

Although we cannot estimate the signal subspace from multiple observations as in the previous section, there is still a way to extract a time-dependent signal subspace from a single observation data. Suppose a sample rate $T_s$, for a single observation data $\bm x = [x(0\cdot T_s),x(1\cdot T_s),\cdots,x((N-1)T_s)]^T$, we can construct its Hankel matrix (also referred to as trajectory matrix), which is
\begin{small}
	\begin{eqnarray}
\bm X = \left[ \begin{array}{cccc}
x(0\cdot T_s) & x(1\cdot T_s) & \cdots & x((K-1)T_s) \\
x(1\cdot T_s) & x(2\cdot T_s) & \cdots & x(KT_s) \\
\vdots & \vdots & \ddots & \vdots \\
x((M-1)T_s) & x(MT_s) & \cdots & x((N-1)T_s)
\end{array}
\right]\label{HankelX}
\end{eqnarray}
\end{small}
where $1 < M < N, K = N-M+1$. The left singular vectors of the Hankel matrix $\bm X$ are known contain
important information about the signal $\bm x$ \cite{jolliffe2005principal}. Therefore the subspace spanned by a subset of these left singular vectors is called "the signal subspace" (generally the left singular vectors corresponding to larger singular values would be chosen). As a matter of fact, this signal subspace extracted from the Hankel matrix has been used to perform noise reduction, signal forecasting, and change point detection, etc \cite{brillinger2001time,golyandina2013singular}. The Hankel matrix technique can be used to analyze a wide variety of signals, like wireless signals, seismologic, meteorological, geophysical time series as well as economic time series. Because no statistical assumption concerning the signal is needed while performing the subspace extraction from Hankel matrices, this methodology is suitable to deal with the fast changing subspace signal and develop our TDOA algorithm. \par
In our setting, at the receiver $i$ ($i=1,2$), given the sampling rate $1/T_s$, the sampled signal vector is $[x_i(0),x(T_s),\cdots,x_i((N-1) \cdot T_s)]^T$ with length $N$, denote the corresponding Hankel matrix as in (\ref{HankelX}) by $\bm X_i\in \mathbb{C}^{M\times K}$, 
$\bm X_i$ can be further written as
\begin{equation}
\bm X_i \approx \sum_{l_i=1}^{L_i}\alpha_{i,l_i}\bm S^{[\tau_{i,l_i}]} + \bm W_i,\label{HankelMulti}
\end{equation}
where $1 < M < N, K = N-M+1$, and
\begin{align}
\bm S^{[\tau_{i,l_i}]} := \hspace{6.2cm} \nonumber \\
\begin{footnotesize}
\left[ \begin{array}{ccc}
s(0\cdot T_s - \tau_{i,l_i}) & \cdots & s((K-1) T_s - \tau_{i,l_i}) \\
s(1 \cdot T_s - \tau_{i,l_i}) &  \cdots & s(K T_s-\tau_{i,l_i}) \\
\vdots & \ddots & \vdots \\
s((M-1) T_s - \tau_{i,l_i}) & \cdots & s((N-1) T_s - \tau_{i,l_i})
\end{array}
\right]	\label{compHankel}
\end{footnotesize}
\end{align}
is the Hankel matrix of the sampled transmitted signal $s(k T_s)$ delayed by $\tau_{i,l_i}$. $\bm W_i$ is the Hankel matrix of noise $w_i(k T_s)$. 

Denote the column vectors of matrix $\bm S^{[\tau_{i,l_i}]}$ by $ \bm s_{i,l_i}^{(1)},\cdots,\bm s_{i,l_i}^{(K)}$, in the following analysis, we just assume these column vectors to have equal length, or instant power, i.e., 
$\|\bm s_{i,l_i}^{(1)}\|_2=\cdots=\|\bm s_{i,l_i}^{(K)}\|_2$.
This assumption is quite general because signals we are interested in can be regarded as stationary so the power is treated as time-invariant during the time of a single observation.
The full algorithm for TDOA estimation of fast changing subspace signal is given in the following.

\subsection{TDOA estimation using VCC function for the slowly changing subspace signals: Algorithm}

\hrule
\par
\hspace{1pt} \par
Algorithm 2. \par \textbf{For $\Delta \tau \in \left((-M+1)T_s, (M-1)T_s\right)$,}
\begin{quote}
	
	1) Construct delayed Hankel matrix
	\begin{footnotesize}
	\begin{eqnarray}
	\bm X_1^{[\Delta \tau]} = \hspace{6.8cm} \nonumber \\
	\left[ \begin{array}{ccc}
	x_1(0\cdot T_s-\Delta \tau) & \cdots & x_1((K-1) T_s-\Delta \tau) \\
	x_1(1\cdot T_s-\Delta \tau)  & \cdots & x_1(K T_s -\Delta \tau) \\
	\vdots & \ddots & \vdots \\
	x_1((M-1)T_s-\Delta \tau) & \cdots & x_1((N-1)T_s-\Delta \tau)
	\end{array}
	\right], \nonumber
	\end{eqnarray}
	\end{footnotesize}
	and non-delayed Hankel matrix
	\begin{footnotesize}	
	\begin{eqnarray}
	\bm X_2 = \hspace{6cm}\nonumber \\
	\left[ \begin{array}{ccc}
	x_2(0\cdot T_s) &  \cdots & x_2((K-1) T_s) \\
	x_2(1\cdot T_s)  & \cdots & x_2(K T_s) \\
	\vdots  & \ddots & \vdots \\
	x_2((M-1) T_s)  & \cdots & x_2((N-1) T_s)
	\end{array}
	\right]. \nonumber
	\end{eqnarray}
	\end{footnotesize}	
	\vspace{0.2cm}\par
	2) Compute singular value decomposition of $\bm X_1^{[\Delta \tau]}$ and $\bm X_2$, then we choose a subset of their left singular vectors, i.e.,
\begin{footnotesize}
		\begin{equation}
	\bm u_{1,1}^{[\Delta \tau]},\cdots,\bm u_{1,K_1}^{[\Delta \tau]},\quad 1 \leq K_1 \leq \min(M,K)\nonumber
	\end{equation}
\end{footnotesize}
	and
	\begin{footnotesize}
		\begin{equation}
	\bm u_{2,1},\cdots,\bm u_{1,K_2},\quad 1\leq K_2 \leq \min(M,K)\nonumber
	\end{equation}
	\end{footnotesize}
	which correspond to the singular values $\sigma_{1,1} \geq \sigma_{1,2},\cdots,\geq \sigma_{1,K_1} >0$ of matrix $\bm X_1^{[\Delta \tau]}$ and singular values $\sigma_{2,1} \geq \sigma_{2,2},\cdots,\geq \sigma_{2,K_2}>0$ of matrix $\bm X_2$. Then the matrices
\begin{footnotesize}
		\begin{equation}
		\bm{ U}_{1}^{[\Delta \tau]}:= [\bm u_{1,1}^{[\Delta \tau]},\cdots,\bm u_{1,K_1}^{[\Delta \tau]}] \in \mathbb{C}^{N \times K_1} \nonumber
	\end{equation}
\end{footnotesize}
	and 
	\begin{footnotesize}
		\begin{equation}
			\bm{ U}_{2} := [\bm u_{2,1},\cdots \bm u_{2,K_2}] \in \mathbb{C}^{N \times K_2} \nonumber
		\end{equation}
	\end{footnotesize}
are basis matrices for the signal subspaces of receiver 1 and 2.
	\vspace{0.1cm}\par
	3)
	Calculate the reciprocal of VCC function:\\
	\begin{equation}
	r_{vol}(\Delta \tau):= 1/\Vol_{K_1+K_2}([\bm{ U}_{1}^{[\Delta \tau]}, \bm{ U}_{2}]),\nonumber
	\end{equation}
\end{quote}
\par  \textbf{Find the peaks of $r_{vol}(\Delta \tau)$ and corresponding $\Delta \tau$.}
\vspace{2pt}
\hrule
\vspace{2pt}
\par 

	Remark 1. Similarly $r_{vol}(\Delta \tau)$ is also a continuous function of variable $\Delta \tau$. And because $\Delta \tau$ can not take continuous values in practice, a similar interpolation step can be used to get the signal with any delay $\Delta \tau$ before we construct the Hankel matrix, simulation will also be provided to show the effect of interpolation.

	Remark 2. There are two important parameters when constructing the Hankel matrix, i.e., the dimensions $M$ and $K$. It is difficult to choose these two dimensions in order to meet different requirements in diverse applications \cite{golyandina2013singular}, so in this paper we just choose these two dimensions empirically based on the experiments and simulations.
	
	Remark 3.  The eigenvector extraction procedure can be regarded as both feature extraction and noise reduction. An important parameter affecting the extraction of signal subspace and calculation of VCC function is the dimension of signal subspaces, i.e., $K_i$. This parameter will be also determined empirically. Actually, in the numerical simulation which will be shown in the next section, $K_i$ is chosen to be $3$ times of $ L_i$.

\subsection{Theoretical Guarantee of Algorithm 2}

\par 
\begin{Therm} \label{volRes2_main}
	Denote the auto-correlation function of transmitted signal $s(k T_s)$ by $R_s(\tau)$,
	consider the Hankel matrices $\bm X_1^{[\Delta \tau]}$ and $\bm X_2$ in Algorithm 2, 
		When for any $1\leq l_1 \leq L_1, 1\leq l_2\leq L_2$, $\Delta \tau\neq \tau_{2,l_2} - \tau_{1,l_1}$, given any parameter
		$\mu$ satisfying
		\begin{equation}
		0 < \mu < \frac{1}{C},
		\end{equation}
		if
		\begin{equation}\label{Hankelmucond0}
		\frac{|R_s(\tau)|}{|R_s(0)|}  \leq \mu, \text{ for } |\tau| \geq \Delta \tau_{\min},
		\end{equation}
		then the volume function in algorithm 2 must satisfy
		\begin{equation}\label{volH0_thm}
		\Vol_{K_1+K_2}([\bm U_1^{[\Delta \tau]}, \bm U_2]) \geq (1-\varepsilon^2)^{K_{\min}/2}
		\end{equation}
		where 
		\begin{equation}
		\Delta \tau_{\min}:= \min \{\Delta\tau_{1},\Delta\tau_{2},\Delta\tau_{1,2}\},
		\end{equation} 
		and
		\begin{eqnarray}
		\Delta\tau_1 &=& \min_{\substack{1\leq l_1 \neq {l_1}' \leq L_1 \\ 1\leq k,k'\leq K}}|\tau_{1,l_1}-\tau_{1,{l_1}'}+(k-k')T_s|, \nonumber \\
		\Delta\tau_2 &=& \min_{\substack{1\leq l_2 \neq {l_2}' \leq L_2 \\ 1\leq k,k'\leq K}}|\tau_{2,l_2}-\tau_{2,{l_2}'}+(k-k')T_s|, \nonumber \\
		\Delta \tau_{1,2} &=& \min_{\substack{ 1\leq l_1 \leq L_1, 1 \leq l_2 \leq L_2 \\ 1\leq k,k' \leq K}}|\tau_{1,l_1}+\Delta \tau-\tau_{2,l_2}+(k-k')T_s|, \nonumber
		\end{eqnarray}
		$K_{\min}:=\min\{K_1,K_2\}$, and the parameter
		\begin{equation}
		\varepsilon=C\cdot\mu,
		\end{equation}
		here 
		\begin{equation}
			C=\frac{K\cdot  \sum_{l_1=1}^{L_1}|\alpha_{1,l_1}| s_{1,l_1}^{[\Delta \tau]} \cdot \sum_{l_2=1}^{L_2}|\alpha_{2,l_2}|s_{2,l_2} }{\sigma_{1,K_1}\sigma_{2,K_2}},
		\end{equation}
		$s_{1,l_1}^{[\Delta \tau]}$ and $s_{2,l_2}$ represent column lengths of Hankel matrices $\bm S^{[\tau_{1,l_1}+\Delta \tau]}$ and $\bm S^{[\tau_{2,l_2}]}$, respectively; $\sigma_{1,K_1}$ and $\sigma_{2,K_2}$ are the $K_1$'th and $K_2$'th largest singular values of matrices $\bm X_1^{[\Delta \tau]}$ and $\bm X_2$.
\end{Therm}
\begin{Therm}\label{volRes2_main2}
		If there exists $l_1^*$ and $l_2^*$, such that $\Delta \tau = \tau_{2,l_2^*}-\tau_{1,l_1^{*}}$, given any parameter $\mu$ satisfying
		\begin{equation}\label{Hankelmucond1mu}
		0 < \mu <
		D/{(L-1)},
		\end{equation}
		if 
		\begin{equation}\label{Hankelmucond1}
		\frac{|R_s(\tau)|}{|R_s(0)|}  \leq \mu, \text{ for } |\tau| \geq \Delta \tau_{\min}^*,
		\end{equation}
		then the volume function will satisfy
		\begin{equation}\label{volH1_thm}
		\Vol_{K_1+K_2}([\bm U_1^{[\Delta d \cdot T_s]}, \bm U_2]) \leq (1-\gamma^2)^{K_{\min}/2},
		\end{equation}	
		meanwhile, the right side of (\ref{volH1_thm}) is less than the right side of (\ref{volH0_thm}). Here
		\begin{equation}
		\Delta \tau_{\min}^*:= \min \{\Delta\tau_{1},\Delta\tau_{2}\},
		\end{equation}
		the parameters
		\begin{equation}\label{ParamD}
		D = 
		\sqrt{\frac{ \sigma_{1,K_1}\sigma_{2,K_2}\cdot A\cdot (L-1)}{  (B+B^*)\cdot K}+\frac{1}{4}}-\frac{1}{2},
		\end{equation}
		and
		\begin{equation}\label{Paramgamma}
		\gamma =  \frac{A}{1+(L-1)\mu} - \frac{B^*\cdot K}{\sigma_{1,K_1}\sigma_{2,K_2}}\cdot\mu,
		\end{equation}
		where $\small L = \min\{L_1,L_2\}$, and
		\begin{eqnarray}
			A &:=& \frac{|\alpha_{1,l_1^*}||\alpha_{1,l_2^*}|}{\sqrt{\sum_{l_1 =1}^{L_1}|\alpha_{1,l_1}|^2\sum_{l_2 =1}^{L_2}|\alpha_{2,l_2} |^2}} \nonumber \\
			B &:=&  \sum_{l_1=1}^{L_1}|\alpha_{1,l_1}| s_{1,l_1}^{[\Delta \tau]}\cdot \sum_{l_2=1}^{L_2}|\alpha_{2,l_2}|s_{2,l_2} \nonumber \\
			B^* &:=& \sum_{l_1=1,l_1\neq l_1^*}^{L_1}|\alpha_{1,l_1}| s_{1,l_1}^{[\Delta \tau]}\cdot \sum_{l_2=1,l_2\neq l_2^*}^{L_2}|{\alpha_{2,l_2}}|s_{2,l_2},\nonumber
		\end{eqnarray}
\end{Therm}
\par Theorem \ref{volRes2_main} and \ref{volRes2_main2} provides a similar theoretical guarantee for the proposed algorithm 2 as Theorem \ref{volRes1_main3} and \ref{volRes1_main2} does. As can be seen, (\ref{volH0_thm}) and (\ref{volH1_thm}) ensure the different values of $\Vol_{K_1+K_2}([\bm U_1^{[\Delta \tau]}, \bm U_2])$ when $\Delta \tau$ as well as $\tau_{1,l_1}$ and $\tau_{2,l_2}$ satisfy condition (\ref{Hankelmucond0}) or (\ref{Hankelmucond1}), which is also related with the autocorrelation function $R_s(\tau)$.
\par However, there is a major difference between Theorem \ref{volRes2_main2} and Theorem \ref{volRes1_main2}. When there exists some $l_1^*$ and $l_2^*$, such that 
\begin{equation}
	\Delta \tau = \tau_{2,l_2^*}-\tau_{1,l_1^{*}},\nonumber
\end{equation}
the VCC function $\Vol_{K_1+K_2}([\bm U_1^{[\Delta \tau]}, \bm U_2])$ won't necessarily be zero, as the result in (\ref{volH1_thm}) shows. As a result, the reciprocal $1/\Vol_{K_1+K_2}([\bm U_1^{[\Delta \tau]}, \bm U_2])$ would show a finite peak when (\ref{Hankelmucond1}) holds. And the parameter $\gamma$ shown in (\ref{Paramgamma}), which is mainly dependent on channel's path gains $\alpha_{1,l_1}$ and $\alpha_{2,l_2}$,
will affect the sharpness of their corresponding TDOA peak at location $\Delta \tau$, then influence the precision of estimation of their TDOAs.
\par Similar to the previous analysis, the conditions (\ref{Hankelmucond0}) and (\ref{Hankelmucond1}) also indicate that, as long as the autocorrelation function $R_s(\tau)$ is "sharp", the performance of the proposed algorithm can be theoretically better. Roughly speaking, when the signal has an extremely sharp autocorrelation function $R_s(\tau)$, all the parameters such as $\mu$ and $\Delta \tau_{\min}$, $\Delta \tau_{\min}^*$ can take a small value, meaning a better TDOA resolution and sharper peaks when $\tau$ is at TDOA positions. This also means that the proposed algorithm prefers signals that has a wide bandwidth.

\section{Numerical simulations}


\subsection{TDOA of slowly changing subspace signals}

\begin{figure}[htbp]
	\centering
	\includegraphics[width=0.48\textwidth]{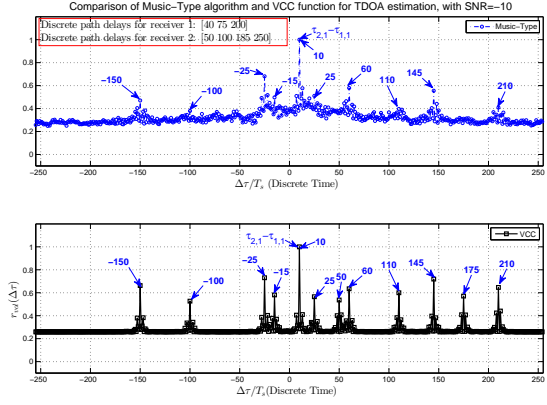}
	\caption{Comparison of Ge's MUSIC-Type algorithm and the VCC algorithm}
	\label{figurevolc}
\end{figure}

Firstly, a demonstration of the TDOA algorithm's output is given by simulation in figure \ref{figurevolc}. In the simulation, a linear frequency modulation (LFM) waveform is chosen as a typical slowly changing subspace signal, which is the most commonly seen waveforms in radar systems. The radar waveform in (\ref{RadarTransmt}) is generated with a sample rate $1MHz$, its length are 2048, and the frequency sweeps linearly from $50kHz$ to $500kHz$. The multipath channel are manually generated, and for convenience, the multipath delays are chosen arbitrarily to be exactly "integer delays", in other words, $\{\tau_{1,l_1}\}_{l_1=1}^{L_1}=\{40 T_s, 75 T_s, 200 T_s\}$ and $\{\tau_{2,l_2}\}_{l_2=1}^{L_2}=\{50 T_s,100 T_s,185 T_s,250 T_s\}$, so that the TDOA can theoretically recovered by testing the VCC function upon integer delays of $\Delta \tau$, i.e., $\Delta \tau = (\cdots,0,1,2,\cdots)\cdot T_s$.
The multiple observations in the form of (\ref{digit_rec_vector_multi}) are directly generated by Monte-Carlo method, in which the channel coefficients $\bm \alpha_{i}^{(j)}=[\alpha_{i,1}^{(j)},\cdots,\alpha_{i,L_i}^{(j)}]$ with respect to different $j$ are generated independently from complex Gaussian distributions in order to simulate the channel fading effect. In addition, the mean value of $|\alpha_{i,1}|$ is greater than the mean value of $|\alpha_{i,l_i}|,l_i>1$, meaning that the direct path has a greater propagation gain than the reflective path. The length $N$ of each observation vector is 512, and totally 512 observation data are generated.

In the simulation, we compare our proposed TDOA algorithm with the publicly known super resolution MUSIC-Type TDOA algorithm proposed by Fengxiang Ge in \cite{ge2007super}, because both algorithms have super resolution and can make use of multiple observation data. Since the simulation focuses on demonstrating the ability of resolving multipath TDOA, we just assume the dimensions of signal subspaces in both algorithms, i.e., the number of channel paths $L_i$, have been accurately estimated. The normalized TDOA estimation results of both algorithms are plotted in figure \ref{figurevolc}, where the signal-to-noise ratio $SNR$, defined as the power ratio of signal and noise, is set to be $-10dB$. According to the simulation setting, there should be peaks at positions where $\Delta \tau = (-150, -100, -25, -15, 10, 25, 50, 60, 110, 145, 175, 210)\cdot T_s$ in the TDOA estimation outputs. The position of these peaks are labeled in the figure. As shown in figure \ref{figurevolc}, when the SNR is low, the MUSIC-Type algorithm fails to reveal most of the peaks of multipath TDOA, but our VCC algorithm can still show clear peaks. 

\begin{figure}[htbp]
	\centering
	\includegraphics[width=0.48\textwidth]{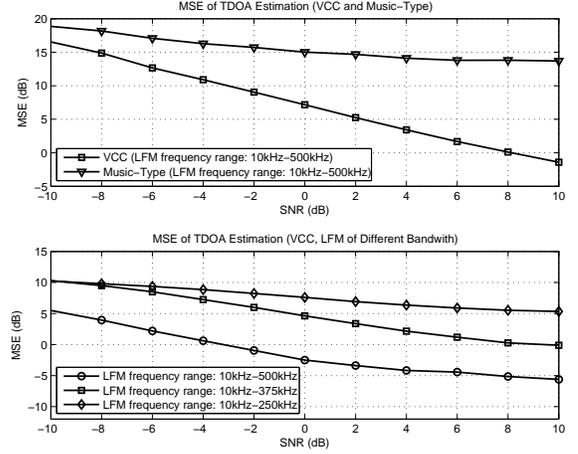}
	\caption{Comparison of the VCC algorithm and MUSIC-Type algorithm (above) and Performance of the VCC algorithm using LFM signal with different frequency range (below)}
	\label{figurevolc_MSE}
\end{figure}

Secondly, the overall performance of both algorithms are given in figure \ref{figurevolc_MSE}. In the simulation, 120 independent trials of both algorithms with arbitrary multipath delays are carried out for different SNR levels. Because the output of these algorithms have different scale, we define the mean square errors (MSE) of TDOA estimation to be $\text{MSE} = \sum_{\Delta \tau} (r(\Delta \tau)-I(\Delta \tau))^2$, where $r(\Delta \tau)$ is the normalized output result shown in figure \ref{figurevolc} of each algorithm, and $I(\Delta \tau)$ is the "standard output vector" which takes value 1 when $\Delta \tau$ is at these multipath TDOA positions and 0 otherwise. The simulation results in figure \ref{figurevolc_MSE} implies that, the proposed VCC algorithm outperforms Ge's Music-Type algorithm at all SNRs; besides, our VCC algorithm will have better performance when source signals have wide bandwidths.
%

\subsection{TDOA of fast changing subspace signals}

In this part of simulation, we chose a set of real-world frequency modulation (FM) broadcast signals as one example of the fast changing subspace signal, to demonstrate the TDOA estimation performance of our proposed method. The FM signals used here are baseband signals transmitted by a real world radio broadcast station gathered from several remote located radio receivers.

\subsubsection{Real world FM signal, when only one single path exists}
\par
In the simulation, the FM signals of a radio station are sampled from two separately located radio receivers, the sample rate of received baseband signals is 256kHz, and the length is 4096.

\begin{figure}[htbp]
	\centering
	\includegraphics[width=0.45\textwidth]{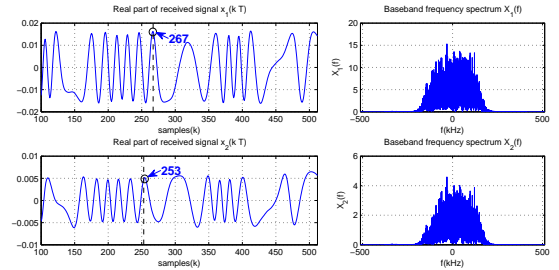}
	\caption{Real world FM signals from two separated receivers}
	\label{figureOrigSignal}
\end{figure}

We firstly increase the original sample rate of the raw signals by a factor of 4 before we use them for TDOA estimation, a part of the waveform in time domain and the frequency spectrum of these two baseband signals are plotted in figure \ref{figureOrigSignal}. From the waveform of both signals, we can see that the corresponding discrete time TDOA is from 14 to 16.

\begin{figure}[htbp]
	\centering
	\includegraphics[width=0.45\textwidth]{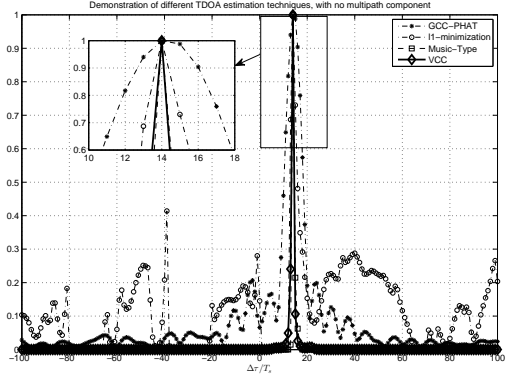}
	\caption{Comparison of different TDOA techniques in a single path channel environment}
	\label{figureMultipleTech}
\end{figure}

In the simulation, we compared our VCC algorithm with the traditional GCC-PHAT method, the high resolution $\ell_1$ regularization algorithm, and also the super resolution MUSIC-Type algorithm by Ge. In the simulation of $\ell_1$ regularization algorithm, the power spectrum of the transmitted signal is required to be known, while the other three algorithms don't use knowledge of the power spectrum. In our algorithm, the parameters $N$, $M$ and $K_i$ are chosen empirically to be $N= 544, M=512,K_1=K_2=3$. The normalized TDOA estimation results of GCC-PHAT, $\ell_1$ regularization, MUSIC-Type as well as VCC algorithm are shown in figure \ref{figureMultipleTech}. It can be seen that in a channel with only a single path, both our proposed VCC algorithm and Ge's MUSIC-Type algorithm outperforms the traditional GCC-PHAT and the $\ell_1$ regularization algorithms; because the latter two methods give a much wider peak, and also reveal too many false peaks except for the real TDOA peak. Although our VCC method and MUSIC-Type algorithm have similar super resolution ability, the computational complexity of our method is much lower.


\subsubsection{Real world FM signal, the multipath channel is manually simulated}
\begin{figure}[htbp]
	\centering
	\includegraphics[width=0.48\textwidth]{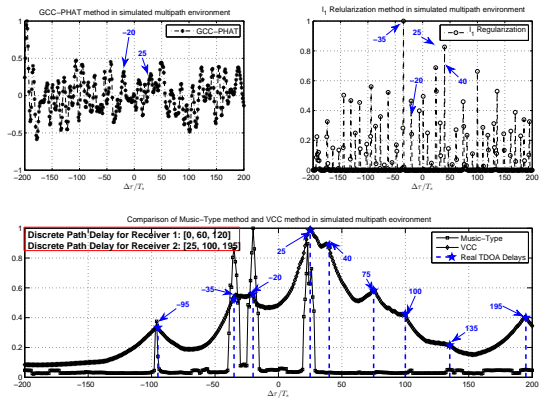}
	\caption{Comparison of different TDOA techniques in a simulated multipath channel environment}
	\label{figureMultipleTech_MP}
\end{figure}

\
In this simulation, the received signals from two receivers are generated as the following expression:
\begin{small}
	\begin{eqnarray}
y_1(k T_s)&=&\alpha_{1,1}s(k T_s)+ \alpha_{1,2}s((k-60) T_s) + \alpha_{1,3}s((k-120) T_s),\nonumber \\
y_2(k T_s)&=&\alpha_{2,1}s((k-25) T_s) + \alpha_{2,2}s((k-100) T_s) \nonumber \\
& &\hspace{3.5cm} + \alpha_{1,3}s((k-195) T_s).\nonumber
\end{eqnarray}
\end{small}
The $s(k T_s)$ here is a real world original FM signal mentioned before, which is also one among the two signals plotted in figure \ref{figureOrigSignal}. The channel coefficients  $\alpha_{i,j},i=1,2,j=1,2,3$ are also generated to simulate a Rician fading channel, among these coefficients the mean value of $|\alpha_{i,1}|$ is greater than that of the other coefficients. In the simulation, the parameters $N$, $M$ and $K_i$ are also chosen empirically to be $N= 896, M= 768,K_1=K_2=9$. The TDOA estimation results of GCC-PHAT, $\ell_1$ regularization, MUSIC-Type and our method are shown in figure \ref{figureMultipleTech_MP}.

As is seen, both Ge's MUSIC-Type algorithm and our VCC algorithm outperforms the other two methods. However, Ge's MUSIC-Type method and our VCC algorithm have their advantages and disadvantages at different aspect. The MUSIC-Type method has a much sharper peak, but fails to resolve every multipath TDOA, and still has some false peaks, while our VCC method may not have such sharp peaks, but successfully reveals every TDOA peak precisely with no false peak. In addition, our VCC algorithm has much lower computational efficiency, because both the MUSIC-Type algorithm and the $\ell_1$ regularization algorithm contain a convex optimization step.
\subsection{Simulations on non-integer multipath delays}
\par
\begin{figure}[htbp]
	\centering
	\includegraphics[width=0.48\textwidth]{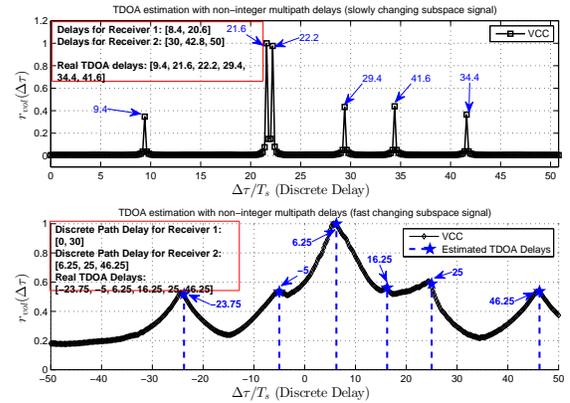}
	\caption{TDOA estimation with non-integer multipath delays}
	\label{figureoffgrid}
\end{figure}

 In the previous simulations, we assume the multipath delays to be exactly on the sampling grid, i.e., $\tau_{i,l_i}/T_s$ takes integer value. But actually, in reality, $\tau_{i,l_i}/T_s$ does not necessary take integer value, as a result, the corresponding TDOA $\Delta \tau$ won't be integer multiple of $T_s$ either. So we're validating the performance of our algorithm when there is non-integer multipath delays. The simulation in figure \ref{figureoffgrid} uses the same signal model as before except for some non-integer delays, it shows that, with a simple interpolation of the received signal before we construct the signal subspace, the non-integer TDOA
delays can still be identified. In the upper figure, we interpolate the sampling rate of received signal $x_i(k T_s)$ by $5$ times, and in the lower figure, the received signal $x_i(k T_s)$ is interpolated by factor $4$. As is shown, the VCC function correctly show peaks at the right TDOA positions. This also suggests that interpolation can be used as a pre-process before the VCC algorithm to improve the precision of TDOA estimation.

\section{Conclusion}

In this paper, a super resolution TDOA estimation technique using the Volume Cross-Correlation function is proposed. This technique firstly estimates the unknown signal subspace from the received signal, and estimate the time difference through the novel VCC function, which calculates the linear dependency of these subspaces. We analyzed the performance of our TDOA estimation algorithm upon two typical categories of signals, i.e., the slowly changing subspace signal and the fast changing subspace signal. Analysis and numerical simulations have demonstrated that our algorithm has excellent capability of super resolution for TDOA estimation in a multipath environment. 

\bibliography{IEEEabrv,VolumeCorrelationTDOA}
\bibliographystyle {IEEEtran}

\appendix
\subsection{\textbf{Proof of Lemma \ref{volRes1_main}, Theorems \ref{volRes1_main3} and \ref{volRes1_main2}:}}
\par 
Several lemmas are needed first:
\begin{Lemma}\label{cohSuff}
	Given a matrix $\bm X = [\bm x_{1},\cdots,\bm x_{L}]\in \mathbb{C}^{N \times L}, L < N$, denote its maximal column correlation (or coherence) by
	\begin{equation}
	\mu := \max_{l\neq l'}\frac{|\langle \bm x_{l},\bm x_{l'} \rangle|}{\|\bm x_{l}\|_2\cdot \|\bm x_{l'}\|_2},
	\end{equation}
	then if
\begin{small}
		\begin{equation}\label{mucond}
	\mu < \frac{1}{L-1},
	\end{equation}
\end{small}
	the matrix $\bm X$ is full rank.
\end{Lemma}
\par
\textbf{Proof of Lemma \ref{cohSuff}:}
This lemma is actually the famous Gershgorin circle theorem, the proof can be found in commonly seen textbooks on matrix analysis, therefore we ignore the proof here.
\begin{Lemma}\label{volRes1}
	Consider the matrices $\bm X_1 = [\bm x_{1,1},\cdots,\bm x_{1,L_1}]\in \mathbb{C}^{N \times L_1}$ and $\bm X_2 = [\bm x_{2,1},\cdots,\bm x_{2,L_2}]\in \mathbb{C}^{N \times L_2}, L_1, L_2 < N$，
	denote the maximum column correlation (also known as coherence) of matrix $[\bm X_1,\bm X_2]$, as well as the maximum column correlation of matrices $\bm X_1$ and $\bm X_2$ by
	\begin{eqnarray}
	\mu_1 &:=& \max_{1\leq l_1 \neq {l_1}'\leq L_1}\frac{|\langle \bm x_{1,l_1},\bm x_{1,{l_1}'}\rangle|}{\|\bm x_{1,l_1}\|_2\cdot\|\bm x_{1,{l_1}'}\|_2}, \label{cohdef2}\\
	\mu_2 &:=& \max_{1\leq l_2 \neq {l_2}'\leq L_2}\frac{|\langle \bm x_{2,l_2},\bm x_{2,{l_2}'}\rangle|}{\|\bm x_{2,l_2}\|_2\cdot\|\bm x_{2,{l_2}'}\|_2},\label{cohdef3}\\
	\mu_0 &:=& \max_{1\leq l_1\leq L_1, 1\leq l_2\leq L_2}\frac{|\langle \bm x_{1,l_1},\bm x_{2,l_2}\rangle|}{\|\bm x_{1,l_1}\|_2\cdot\|\bm x_{2,l_2}\|_2},\label{cohdef1},
	\end{eqnarray}
	taking $\mu = \max\{\mu_0,\mu_1,\mu_2\}$, then if
	\begin{equation}\label{Volcond}
	\mu \leq \frac{1}{L_1+L_2-1},
	\end{equation}
	we have
	\begin{equation}
	\Vol_{L_1+L_2}([\bm U_{1}, \bm U_{2}])\geq (1-\varepsilon)^{L/2}, \nonumber
	\end{equation}
	where the matrices $\bm U_{1}, \bm U_{2}$ are orthogonal bases for subspaces $\Span(\bm X_1)$ and $\Span(\bm {X}_2)$, and $L=\min\{L_1,L_2\}$,
\begin{small}
		\begin{equation}
	\varepsilon = \frac{L_1L_2\cdot \mu^2}{[1-(L_1-1)\mu][1-(L_2-1)\mu]}.\nonumber
	\end{equation}
\end{small}	
\end{Lemma}
\textbf{Proof of Lemma \ref{volRes1}:}\par
	Similar to the previous proof, we use the column-normalized versions of matrices $\bm X_1$ and $\bm X_2$, which are denoted by $\tilde{\bm X}_1=[\tilde{\bm x}_{1,1},\cdots,\tilde{\bm x}_{1,L_1}],\tilde{\bm X}_2=[\tilde{\bm x}_{2,1},\cdots,\tilde{\bm x}_{2,L_2}]$ in the following proof. Therefore we have $\tilde{\bm x}_{1,l_1} = \bm x_{1,l_1}/\|\bm x_{1,l_1}\|_2,l_1=1,\cdots,L_1$, and $\tilde{\bm x}_{2,l_2} = \bm x_{2,l_2}/\|\bm x_{2,l_2}\|_2,l_2=1,\cdots,L_2$. It is easy to verify that $\Span(\tilde{\bm X}_1) = \Span(\bm X_1)$ and $\Span(\tilde{\bm X}_2) = \Span(\bm X_2).$
	
	According to the relation of volume and principal angles, the volume function $\Vol_{L_1+L_2}([\bm U_{1}, \bm U_{2}])$ satisfies
	\begin{equation}
	\Vol_{L_1+L_2}([\bm U_{1}, \bm U_{2}]) = \prod_{i=1}^{L}\sin \theta_i(\Span(\tilde{\bm X}_1),\Span(\tilde{\bm X}_2)),\nonumber
	\end{equation}
	where $L=\min\{L_1,L_2\}$, and $\theta_i(\Span(\tilde{\bm X}_1),\Span(\tilde{\bm X}_2))$ are principal angles of subspaces $\Span(\tilde{\bm X}_1)$ and $\Span(\tilde{\bm X}_2)$.
	From the definition of principal angles, it can be derived that, given any principal angle $\theta_i(\Span(\tilde{\bm X}_1),\Span(\tilde{\bm X}_2)),i=1,\cdots,L$, there must exist a pair of vectors $\bm u_i$ and $\bm v_i$, which satisfy
	\begin{eqnarray}
	\bm u_i \in \Span(\tilde{\bm X}_1),\|\bm u_i\|_2 = 1, \\
	\bm v_i \in \Span(\tilde{\bm X}_2), \|\bm v_i\|_2 = 1, 
	\end{eqnarray}
	and
	\begin{equation}
	\cos \theta_i(\Span(\tilde{\bm X}_1),\Span(\tilde{\bm X}_2))= |\langle \bm u_i, \bm v_i \rangle|.
	\end{equation}
	Then according to Lemma \ref{cohSuff}, the condition in (\ref{Volcond}) implies the full-rankness of matrix $[\tilde{\bm X}_1,\tilde{\bm X}_2]$, which also means that $\tilde{\bm X}_1$ and $\tilde{\bm X}_2$ are full rank; as a result, there must exist a series of coefficients $a_{1,1},\cdots, a_{1,L_1}$ and $a_{2,1},\cdots,a_{2,L_2}$ that are not all zero, such that
\begin{small}
		\begin{eqnarray}
	\bm u_i = \sum_{l_1=1}^{L_1} a_{1,l_1} \tilde{\bm x}_{1,l_1},\
	\bm v_i = \sum_{l_2=1}^{L_2} a_{2,l_2} \tilde{\bm x}_{2,l_2},
	\end{eqnarray}
\end{small}
	then we have
	\begin{eqnarray}
	|\langle \bm u_i, \bm v_i \rangle| 
	&\leq& \sum_{l_1=1}^{L_1}\sum_{l_2=1}^{L_2}|a_{1,l_1}a_{2,l_2}| \cdot \mu \nonumber \\
	& \leq & \sqrt{L_1\cdot \sum_{l_1=1}^{L_1}|a_{1,l_1}|^2}\sqrt{L_2\cdot \sum_{l_2=1}^{L_2}|a_{2,l_2}|^2} \cdot \mu,\label{principalIneq}
	\end{eqnarray}
	the last inequality is derived from the Cauchy-Schwarz inequality. 
	\par Because $\|\bm u_i\|_2  = 1$, we also have
	\begin{eqnarray}
	1=\|\bm u_i\|_2^2 
	&\geq&  \sum_{l_1=1}^{L_1} |a_{1,l_1}|^2 - \sum_{l_1\neq {l_1}'} |a_{1,l_1} a_{1,{l_1}'}| \mu \nonumber \\
	&\geq& \sum_{l_1=1}^{L_1} |a_{1,l_1}|^2 - (L_1-1)\sum_{l_1=1}^{L_1} |a_{1,l_1}|^2\cdot \mu,\label{lengthconst}
	\end{eqnarray}
	the last inequality is also derived from the Cauchy-Schwarz inequality. Then we can know from (\ref{lengthconst}) that,
	\begin{equation}\label{coeff1}
	\sum_{l_1=1}^{L_1} |a_{1,l_1}|^2 \leq \frac{1}{1-(L_1-1)\mu},	
	\end{equation}
	similarly we also have
	\begin{equation}\label{coeff2}
	\sum_{l_2=1}^{L_2} |a_{2,l_2}|^2 \leq \frac{1}{1-(L_2-1)\mu},
	\end{equation}
	combining (\ref{coeff1}) and (\ref{coeff2}) with (\ref{principalIneq}), we have
	\begin{small}
		\begin{eqnarray}
	\cos \theta_i(\Span(\tilde{\bm X}_1),\Span(\tilde{\bm X}_2))= |\langle \bm u_i, \bm v_i \rangle|  \hspace{1.5cm}\nonumber\\
	 \leq \sqrt{\frac{L_1\cdot L_2\cdot \mu^2}{[1-(L_1-1)\mu][1-(L_2-1)\mu]}},
	\end{eqnarray}
	\end{small}
	or in another word,
	\begin{small}
		\begin{eqnarray}
	\sin \theta_i(\Span(\tilde{\bm X}_1),\Span(\tilde{\bm X}_2)) \geq \hspace{3cm} \nonumber \\ \sqrt{1-\frac{L_1\cdot L_2\cdot \mu^2}{[1-(L_1-1)\mu][1-(L_2-1)\mu]}},
	\end{eqnarray}
	\end{small}
	holds for every $i,i=1,\cdots,L$. If we let $\varepsilon = \frac{L_1\cdot L_2\cdot \mu^2}{[1-(L_1-1)\mu][1-(L_2-1)\mu]}$, Lemma \ref{volRes1} is now proved.
\par \textbf{Proof of Lemma \ref{volRes1_main}, Theorems \ref{volRes1_main3} and \ref{volRes1_main2}:}
\par First, Lemma \ref{volRes1_main} is derived based on Lemma \ref{cohSuff}. Because the matrices $\bm G_1^{[\Delta d\cdot T_s]}$ in (\ref{subbasis1}) and $\bm G_2$ in (\ref{subbasis2}) are actually composed of different delayed versions of the waveform signal $g(k T_s)$, so (\ref{mucond}) is actually constraining the auto-correlation function of $g(k T_s)$. Without loss of generality, we assume the signal $g(k T_s)$ to be wide sense stationary, then we can approximately have
\begin{footnotesize}
	\begin{eqnarray}
		\frac{|\langle \bm g_{i,l_i}, \bm g_{i,{l_i}'} \rangle |}{\|\bm g_{i,l_i}\|_2\|\bm g_{i,{l_i}'}\|_2} &=& \frac{|R_s(\tau_{i,l_i}-\tau_{i,{l_i}'})|}{|R_s(0)|},\text{ for }  l_i \neq l_i'\nonumber
	\end{eqnarray}
\end{footnotesize}
thus (\ref{cohdef2}) and (\ref{cohdef3}) means
\begin{equation}
\max_{1\leq l_i\neq {l_i}' \leq L_i} |R_g(\tau_{i,l_i}-\tau_{i,{l_i}'})| \leq \frac{|R_g(0)|}{L_1+L_2-1}, \ i=1,2, \label{ACFcond} 
\end{equation}
because we are deriving a sufficient condition, the condition in (\ref{ACFconst1}) will surely be sufficient to ensure (\ref{ACFcond}).
the proof of Lemma \ref{volRes1_main} is completed.
\par As for Theorem \ref{volRes1_main3}, the proof is based on Lemma \ref{volRes1}, simililarly, (\ref{cohdef1}) is equivalently written as:
\begin{equation}
	\max_{1\leq l_1 \leq L_1, 1\leq l_2 \leq L_2} |R_g((\tau_{1,l_1}+\Delta \tau-\tau_{2,l_2})T_s)| \leq \frac{|R_g(0)|}{L_1+L_2-1}, \nonumber
\end{equation}
then similarly the condition in (\ref{ACFconst2}) is sufficient.
\par Theorem \ref{volRes1_main2} is obvious and no need to prove.

\subsection{\textbf{Proof of Theorems \ref{volRes2_main} and \ref{volRes2_main2}:}}
\par Before proving Theorems \ref{volRes2_main} and \ref{volRes2_main2}, the following two lemmas are firstly given as intermediate results.
\begin{Lemma}\label{HankelH0}
	Consider the linear combinations of several matrices:
\begin{small}
	\begin{equation}
	\bm H_1 = \sum_{l_1=1}^{L_1}\alpha_{1,l_1}\bm H_{1,l_1} = \left[\sum_{l_1=1}^{L_1} \alpha_{1,l_1}\bm h_{1,l_1}^{(1)},\cdots,\sum_{l_1=1}^{L_1}\alpha_{1,l_1}\bm h_{1,l_1}^{(K)}\right],\label{Hankelcomb1}
	\end{equation}
\end{small}
	and
	\begin{small}
		\begin{equation}
	\bm H_2 = \sum_{l_2=1}^{L_2}\alpha_{2,l_2}\bm H_{2,l_2} = \left[\sum_{l_2=1}^{L_2}\alpha_{2,l_2}\bm h_{2,l_2}^{(1)},\cdots,\sum_{l_2=1}^{L_2}\alpha_{2,l_2}\bm h_{2,l_2}^{(K)}\right],\label{Hankelcomb2}
	\end{equation}	
	\end{small}
	with $\bm H_1  \in \mathbb{C}^{M \times K}$ and $\bm H_2 = \in\mathbb{C}^{M \times K}$, and $\|\bm h_{1,l_1}^{(1)}\|_2=\cdots=\|\bm h_{1,l_1}^{(K)}\|_2$, $\|\bm h_{2,l_2}^{(1)}\|_2=\cdots=\|\bm h_{2,l_2}^{(K)}\|_2$; we let
	\begin{eqnarray}
	\mu_1 &=& \max_{\substack{1\leq k,k'\leq K\\ 1\leq l_1 \neq {l_1}' \leq L_1 }}\frac{|\langle \bm h_{1,l_1}^{(k)}, \bm h_{1,{l_1}'}^{(k')} \rangle |}{\|\bm h_{1,l_1}^{(k)}\|_2\|\bm h_{1,{l_1}'}^{(k')}\|_2} \label{Hankelcoh01}\\
	\mu_2 &=& \max_{\substack{1\leq k,k'\leq K \\ 1\leq l_2 \neq {l_2}' \leq L_2 }}\frac{|\langle \bm h_{2,l_2}^{(k)}, \bm h_{2,{l_2}'}^{(k')} \rangle |}{\|\bm h_{2,l_2}^{(k)}\|_2\|\bm h_{2,{l_2}'}^{(k')}\|_2}\label{Hankelcoh02}
	\end{eqnarray}
	and let
	\begin{equation}
		\mu_0 = \max_{\substack{1\leq k,k' \leq K \\ 1\leq l_1 \leq L_1,1 \leq l_2 \leq L_2}}\frac{|\langle\bm h_{1,l_1}^{(k)}, \bm h_{2,l_2}^{(k')} \rangle|}{\|\bm h_{1,l_1}^{(k)}\|_2\|\bm h_{2,l_2}^{(k')}\|_2}. \label{Hankelcoh00}
	\end{equation}
	Taking
	\begin{equation}
	\mu = \max\{\mu_0,\mu_1,\mu_2\},
	\end{equation}
	then we have the same result as in (\ref{Hankelmucond0}) and (\ref{volH0_thm}).
	
\end{Lemma}
\begin{Lemma}\label{HankelH1}
	Consider the matrices $\bm H_1$ and $\bm H_2$ in (\ref{Hankelcomb1}) and (\ref{Hankelcomb2}), if there exists $1\leq l_1^*\leq L_1$ and $1\leq l_2^* \leq L_2$, such that $\bm H_{1,l_1^*} = \bm H_{2,l_2^*}$; we let
	\begin{eqnarray}
	\mu_0 &=& \max_{\stackrel{1\leq k,k' \leq K}{1\leq l \leq L_1,l \neq l_1^*,1 \leq l_1 \leq L_2,l_2\neq l_2^*}}\frac{|\langle\bm h_{1,l_1}^{(k)}, \bm h_{2,l_2}^{(k')} \rangle|}{\|\bm h_{1,l_1}^{(k)}\|_2\|\bm h_{2,l_2}^{(k')}\|_2} \label{Hankelcoh10}
	\end{eqnarray}
	$L: = \min\{L_1, L_2\}$, and let
	\begin{equation}
	\mu = \max\{\mu_0,\mu_1,\mu_2\}
	\end{equation}
	then we have the same results as in (\ref{Hankelmucond1}) and (\ref{volH1_thm}).
\end{Lemma}
 \par
\textbf{Proof of Lemma \ref{HankelH0}:}
	For simplicity, we take the matrix $\bm H_1$ for example. Before we start the proof, for convenience, we will use the column-normalized  matrices of $\bm H_{1,l_1}$ as our main target, namely, we take
	$\hat{\bm H}_{1,l_1} = [\hat{\bm h}_{1,l_1}^{(1)},\cdots,\hat{\bm h}_{1,l_1}^{(K)}]=\left[\bm h_{1,l_1}^{(1)}/\|\bm h_{1,l_1}^{(1)}\|_2,\cdots,\bm h_{1,l_1}^{(K)}/\|\bm h_{1,l_1}^{(K)}\|_2\right]$.
	Because we have assumed $\|\bm h_{1,l_1}^{(1)}\|_2=\cdots=\|\bm h_{1,l_1}^{(K)}\|_2$, then the matrices $\bm H_1$ can be equivalently written into
	\begin{footnotesize}
		\begin{eqnarray}
	\bm H_1 = \left[\sum_{l_1=1}^{L_1}\hat \alpha_{1,l_1}\hat{\bm h}_{1,l_1}^{(1)},\cdots,\sum_{l_1=1}^{L_1}\hat \alpha_{1,l_1}\hat{\bm h}_{1,l_1}^{(K)}\right] 
	\end{eqnarray}
	\end{footnotesize}
	where $\hat \alpha_{1,l_1} = \alpha_{1,l_1}\cdot\|\bm h_{1,l_1}^{(k)}\|_2, k=1,\cdots, K$. 
	\par According to the definition of singular value decomposition, we have
\begin{footnotesize}
		\begin{eqnarray}
	\bm H_1 = \sum_{i=1}^{\min\{M,K\}}\sigma_{1,i}\bm u_{1,i}\bm v_{1,i}^H, 
	\end{eqnarray}
\end{footnotesize}
	Then the left singular vectors of matrix $\bm H_1$ (or $\bm H_2$), corresponding to its $K_1$ largest singular values, namely those $\bm u_{1,1},\cdots,\bm u_{1,K_1}$, can be written as
	\begin{eqnarray}
	\bm u_{1,r} =\frac{1}{\sigma_{1,r}} \bm H_1 \bm v_{1,r}, \quad r = 1,\cdots,K_1,
	\end{eqnarray}
	As a matter of fact, these vectors are actually regarded as the basis for each receiver's signal subspaces, and the basis matrices are 
	\begin{small}
		\begin{eqnarray}
	\bm U_1 = [\bm u_{1,1},\cdots,\bm u_{1,K_1}], \ 
	\bm U_2 = [\bm u_{1,2},\cdots,\bm u_{1,K_2}] 
	\end{eqnarray}
	\end{small}
	On the other hand, according to the definition of principal angles, take $K_{\min} = \min\{K_1,K_2\}$, then for every principal angle of subspaces $\Span(\bm U_1)$ and $\Span(\bm U_2)$, i.e., $\theta_i(\Span(\bm U_1),\Span(\bm U_2)),i = 1,\cdots,K_{\min}$, there must exist a pair of vectors $\tilde{\bm u}_{1,i}$ and $\tilde{\bm u}_{2,i}$, satisfying
	\begin{eqnarray}
	\tilde{\bm u}_{1,i} \in \Span(\bm U_1),\|\tilde{\bm u}_{1,i}\|_2 = 1, \nonumber\\
	\tilde{\bm u}_{2,i} \in \Span(\bm U_2), \|\tilde{\bm u}_{2,i}\|_2 = 1, \nonumber 
	\end{eqnarray}
	such that
	\begin{equation}
	\cos \theta_i(\Span(\bm U_1),\Span(\bm U_2))= |\langle \tilde{\bm u}_{1,i}, \tilde{\bm u}_{2,i} \rangle|.\nonumber
	\end{equation}
	As a matter of fact, these pair of vectors whose angles are principal angles, i.e., $\tilde{\bm u}_{1,i}$ and $\tilde{\bm u}_{2,i}$ for $i = 1,\cdots, K_{\min}$, are actually the left and right singular vectors of matrix $\bm U_1^H\bm U_2$. Therefore, they can be orthogonally represented by the orthogonal basis matrices $\bm U_1$ and $\bm U_2$, which means, there exist a series of coefficients $q_{1,i}^{(1)},\cdots,q_{1,i}^{(K_1)}$ and $q_{2,i}^{(1)},\cdots,q_{2,i}^{(K_2)}$ that are not all zero, such that
	\begin{eqnarray}
	\tilde{\bm u}_{1,i} = \sum_{r=1}^{K_1}q_{1,i}^{(r)}\bm u_{1,r},\ 
	\tilde{\bm u}_{2,i} = \sum_{r'=1}^{K_2}q_{2,i}^{(r')} \bm u_{2,r'}, \nonumber
	\end{eqnarray}
	also $\sum_{r=1}^{K_1}|q_{1,i}^{(r)}|^2=\sum_{r'=1}^{K_2}|q_{2,i}^{(r')}|^2=1$. If we denote $\bm q_{1,i}:=[q_{1,i}^{(1)},\cdots,q_{1,i}^{(K_1)}]^T$,
	then we have
	\begin{equation}
	\tilde{\bm u}_{1,i} = \bm H_1\cdot [\bm v_{1,1},\cdots,\bm v_{1,K_1}]\cdot \text{diag}(\sigma_{1,1}^{-1},\cdots,\sigma_{1,K_1}^{-1})\cdot \bm q_{1,i}, \label{principal01} 
	\end{equation}
	letting $\tilde{\bm v}_{1,i} = [\bm v_{1,1},\cdots,\bm v_{1,r_1}]\cdot \text{diag}(\sigma_{1,1}^{-1},\cdots,\sigma_{1,K_1}^{-1})\cdot \bm q_{1,i}$, 
	then (\ref{principal01}) are equivalently
	\begin{eqnarray}
	\tilde{\bm u}_{1,i} = \bm H_1\cdot \tilde{\bm v}_{1,i} = \sum_{k=1}^{K}\tilde{v}_{1,i}^{(k)}\sum_{l_1=1}^{L_1}\hat \alpha_{1,l_1}\hat{\bm h}_{1,l_1}^{(k)}, \label{principal01_main} 
	\end{eqnarray}
	where $\tilde{\bm v}_{1,i} = [\tilde{v}_{1,i}^{(1)},\cdots,\tilde{v}_{1,i}^{(K)}]^T$. It is not hard to prove that
	\begin{eqnarray}
	\|\tilde{\bm v}_{1,i}\|_2 = \|\text{diag}(\sigma_{1,1}^{-1},\cdots,\sigma_{1,K_1}^{-1})\cdot \bm q_{1,i}\|_2 \leq \sigma_{1,K_1}^{-1}. \label{length1}
	\end{eqnarray}
	Similarly, if we denote $\bm q_{2,i}:=[q_{1,i}^{(1)},\cdots,q_{1,i}^{(K_2)}]^T$, we have similar results:
	\begin{eqnarray}
		\tilde{\bm u}_{2,i} &=& \bm H_2\cdot \tilde{\bm v}_{2,i} = \sum_{k=1}^{K}\tilde{v}_{2,i}^{(k)}\sum_{l_2=1}^{L_2}\hat \alpha_{2,l_2}\hat{\bm h}_{2,l_2}^{(k)},  \label{principal02_main} \\
		\|\tilde{\bm v}_{2,i}\|_2 &=& \|\text{diag}(\sigma_{2,1}^{-1},\cdots,\sigma_{2,K_1}^{-1})\cdot \bm q_{2,i}\|_2 \leq \sigma_{2,K_2}^{-1}. \label{length2}
	\end{eqnarray}
	According to the previous analysis, we have
	\begin{footnotesize}
		\begin{eqnarray}
	\lefteqn{|\langle\tilde{\bm u}_{1,i}, \tilde{\bm u}_{2,i}\rangle|}&& \nonumber \\&=& |\sum_{k=1}^{K}\sum_{k'=1}^{K}\tilde{v}_{1,i}^{(k)}\overline{\tilde{v}_{2,i}^{(k')}}\langle\sum_{l_1=1}^{L_1}\hat\alpha_{1,l_1}\hat{\bm h}_{1,l_1}^{(k)}, \sum_{l_2=1}^{L_2}\hat \alpha_{2,l_2}\hat{\bm h}_{2,l_2}^{(k')}\rangle|\hspace{0.5cm} \label{innerprod} \\
	&\leq& \sum_{l=1}^{L_1}|\hat \alpha_{1,l}|\cdot \sum_{k=1}^{K}|\tilde{v}_{1,i}^{(k)}|\cdot \sum_{l_2=1}^{L_2}|\hat \alpha_{2,l_2}|\cdot \sum_{k'=1}^{K}|\tilde{v}_{2,i}^{(k')}|\cdot \mu \nonumber \\
	&\leq & \sum_{l=1}^{L_1}|\hat \alpha_{1,l}|\sqrt{K\cdot \sum_{k=1}^{K}|\tilde{v}_{1,i}^{(k)}|^2}\cdot \sum_{l_2=1}^{L_2}|\hat \alpha_{2,l_2}|\sqrt{K\cdot \sum_{k'=1}^{K}|\tilde{v}_{2,i}^{(k')}|^2}\cdot \mu \hspace{0.5cm}\label{innerprodlbd} \\
	&\leq & \sum_{l_1=1}^{L_1}\left|\hat\alpha_{1,l_1}\right|\cdot  \sum_{l_2=1}^{L_2}\left|\hat \alpha_{2,l_2}\right| \cdot\frac{K\cdot \mu}{\sigma_{1,K_1}\sigma_{2,K_2}} 	\label{innerprodlbd2}
	\end{eqnarray}
	\end{footnotesize}
	holds for every $i = 1,\cdots,K_{\min}$. The inequality in (\ref{innerprodlbd}) is based on the Cauchy-Schwarz inequality, and the inequality in (\ref{innerprodlbd2}) is based on (\ref{length1}) and (\ref{length2}).
	If we take $C_1 = \sum_{l_1=1}^{L_1}\left|\hat\alpha_{1,l_1}\right|, C_2 = \sum_{l_2=1}^{L_2}\left|\hat \alpha_{2,l_2}\right|$, Lemma \ref{HankelH0} is now proved.
\textbf{Proof of Lemma \ref{HankelH1}:}
	\par Similar to the proof of Lemma \ref{HankelH0}, we use (\ref{principal01_main}) and (\ref{principal02_main}) to prove Lemma \ref{HankelH1}. According to (\ref{innerprod}), we have
\begin{footnotesize}
		\begin{eqnarray}
	\lefteqn{|\langle\tilde{\bm u}_{1,i}, \tilde{\bm u}_{2,i}\rangle| = |\sum_{k=1}^{K}\sum_{k'=1}^{K}\tilde{v}_{1,i}^{(k)}\overline{\tilde{v}_{2,i}^{(k')}}\langle\sum_{l=1}^{L_1}\hat \alpha_{1,l_1}\hat{\bm h}_{1,l_1}^{(k)}, \sum_{l_2=1}^{L_2}\hat \alpha_{2,l_2}\hat{\bm h}_{2,l_2}^{(k')}\rangle|} && \nonumber \\
	&\geq & |\hat\alpha_{1,l_1^*}\hat\alpha_{2,l_2^*}|\cdot\sum_{k=1}^{K}|\tilde{v}_{1,i}^{(k)}|\cdot\sum_{k'=1}^{K}|\tilde{v}_{2,i}^{(k')}|-\nonumber \\
	& &
	\qquad\sum_{l_1\neq l_1^*}^{L_1}\sum_{l_2\neq l_2^*}^{L_2}|\hat\alpha_{1,l_1}|\cdot |{\hat\alpha_{2,l_2}}|\cdot \sum_{k=1}^{K}\sum_{k'=1}^{K}| \tilde{v}_{1,i}^{(k)}|\cdot|\tilde{v}_{2,i}^{(k')}|\cdot\mu	\nonumber \\
	&\geq&  |\hat\alpha_{1,l_1^*}\hat\alpha_{2,l_2^*}|\cdot\sum_{k=1}^{K}|\tilde{v}_{1,i}^{(k)}|\cdot\sum_{k'=1}^{K}|\tilde{v}_{2,i}^{(k')}|-\nonumber \\
	& &
	\qquad\sum_{l_1\neq l_1^*}^{L_1}\sum_{l_2\neq l_2^*}^{L_2}|\hat\alpha_{1,l_1}|\cdot |{\hat\alpha_{2,l_2}}|\cdot \frac{K}{\sigma_{1,K_1}\sigma_{2,K_2}}\cdot\mu	\label{innerprodH1}
	\end{eqnarray}
\end{footnotesize}
	Then, the constraint $\|\tilde{\bm u}_{1,i}\|_2=1$ is considered, we can prove
\begin{footnotesize}
		\begin{eqnarray}
	\lefteqn{\|\tilde{\bm u}_{1,i}\|_2^2\leq \sum_{l_1=1}^{L_1}\|\sum_{k=1}^{K}\hat\alpha_{1,l_1}\tilde{v}_{1,i}^{(k)}\hat{\bm h}_{1,l_1}^{(k)}\|_2^2} & & \nonumber \\
	& & +  \left[\left|\sum_{l_1 =1}^{L_1}\hat\alpha_{1,l_1}\right|^2\cdot \left|\sum_{k=1}^{K}\tilde{v}_{1,i}^{(k)}\right|^2 - \sum_{l_1 =1}^{L_1}|\hat\alpha_{1,l_1}|^2\left|\sum_{k=1}^{K}\tilde{v}_{1,i}^{(k)}\right|^2\right] \mu\hspace{0.5cm}\label{l2const}
	\end{eqnarray}
\end{footnotesize}
	because the geometrical average is greater than the arithmetic average, we have
	\begin{footnotesize}
		\begin{equation}
	\frac{\sum_{l_1 =1}^{L_1}\left|\hat\alpha_{1,l_1}\right|  }{L_1} \leq \sqrt{ \frac{\sum_{l_1 =1}^{L_1}\left|\hat\alpha_{1,l_1}\right|^2  }{L_1}},
	\end{equation}
	\end{footnotesize}
	therefore (\ref{l2const}) becomes
\begin{footnotesize}
		\begin{align}
	\|\tilde{\bm u}_{1,i}\|_2^2 \leq \sum_{l_1=1}^{L_1}\|\sum_{k=1}^{K}\alpha_{1,l_1}^{(k)}\tilde{v}_{1,i}^{(k)}\hat{\bm h}_{1,l_1}^{(k)}\|_2^2 +  \hspace{2cm}\nonumber \\ \sum_{l_1 =1}^{L_1}|\hat\alpha_{1,l_1}|^2\left|\sum_{k=1}^{K}\tilde{v}_{1,i}^{(k)}\right|^2(L_1-1) \mu,
	\end{align}
\end{footnotesize}
	because
	\begin{eqnarray}
	\|\sum_{k=1}^{K}\hat\alpha_{1,l_1}\tilde{v}_{1,i}^{(k)}\hat{\bm h}_{1,l_1}^{(k)}\|_2^2
	&\leq& |\hat\alpha_{1,l_1}|^2\cdot|\sum_{k=1}^{K}\tilde{v}_{1,i}^{(k)}|^2
	\end{eqnarray}
	(\ref{l2const}) becomes
\begin{small}
		\begin{equation}
	\|\tilde{\bm u}_{1,i}\|_2^2 \leq \sum_{l_1 =1}^{L_1}|\hat\alpha_{1,l_1}|^2\left|\sum_{k=1}^{K}\tilde{v}_{1,i}^{(k)}\right|^2 \left[ 1+  (L_1-1)\right] \mu,
	\end{equation}
\end{small}
	therefore we have
	\begin{equation}
	\left|\sum_{k=1}^{K}\tilde{v}_{1,i}^{(k)}\right|^2 \geq \frac{1}{\sum_{l =1}^{L_1}|\hat\alpha_{1,l}|^2\left[1+(L_1-1)\mu\right]}.
	\end{equation}
	Using the same technique, we can derive
	\begin{equation}
	\left|\sum_{k'=1}^{K}\tilde{v}_{2,i}^{(k')}\right|^2 \geq \frac{1}{\sum_{l_2 =1}^{L_1}|\hat\alpha_{2,l_2}|^2\left[1+(L_2-1)\mu\right]}.
	\end{equation}	
	Combined with (\ref{innerprodH1}) we have
	\begin{footnotesize}
		\begin{eqnarray}
	\lefteqn{|\langle\tilde{\bm u}_{1,i}, \tilde{\bm u}_{2,i}\rangle| \geq} &&\nonumber \\
	& &\frac{|\hat\alpha_{1,l_1^*}\hat\alpha_{2,l_2^*}|}{\sqrt{\sum_{l_1 =1}^{L_1}|\hat\alpha_{1,l_1}|^2\sum_{l_2 =1}^{L_2}|\hat\alpha_{2,l_2}|^2}}\cdot \sqrt{\frac{1}{\left[1+(L_1-1)\mu\right]\left[1+(L_2-1)\mu\right]}} \nonumber \\
	& &\qquad -\sum_{l_1=1,l_1\neq l_1^*}^{L_1}|\hat\alpha_{1,l_1}|\sum_{l_2=1,l_2\neq l_2^*}^{L_2}|{\hat\alpha_{2,l_2}}|\cdot \frac{K}{\sigma_{1,K_1}\sigma_{2,K_2}}\cdot\mu\label{innerpricp}
	\end{eqnarray}
	\end{footnotesize}
	Now (\ref{volH1_thm}) is proved.
We denote
\begin{small}
			\begin{eqnarray}
		A &:=& \frac{|\alpha_{1,l_1^*}||\alpha_{1,l_2^*}|}{\sqrt{\sum_{l_1 =1}^{L_1}|\alpha_{1,l_1}|^2\sum_{l_2 =1}^{L_2}|\alpha_{2,l_2} |^2}} \nonumber \\
		B &:=&  \sum_{l_1=1}^{L_1}|\alpha_{1,l_1}| \|\bm h_{1,l_1}^{(k)}\|_2\cdot \sum_{l_2=1}^{L_2}|\alpha_{2,l_2}|\|\bm h_{2,l_2}^{(k)}\|_2 \nonumber \\
		B^* &:=& \sum_{l_1=1,l_1\neq l_1^*}^{L_1}|\alpha_{1,l_1}| \|\bm h_{1,l_1}^{(k)}\|_2\cdot \sum_{l_2=1,l_2\neq l_2^*}^{L_2}|{\alpha_{2,l_2}}|\|\bm h_{2,l_2}^{(k)}\|_2,\nonumber
		\end{eqnarray}
\end{small}
	Sufficiently, if we let
	the lower bound of (\ref{innerpricp}) be greater than the upper bound of (\ref{innerprodlbd2}), then the right side of (\ref{volH1_thm}) is less than the right side of (\ref{volH0_thm}), the corresponding condition on $\mu$ is thus $\footnotesize \mu \leq \sqrt{\frac{ \sigma_{1,K_1}\sigma_{2,K_2}\cdot A\cdot (L-1)}{  (B+B^*)\cdot K}+\frac{1}{4}}-\frac{1}{2}$ as in (\ref{ParamD}). The lemma is proved.

\textbf{Proof of  Theorems \ref{volRes2_main} and \ref{volRes2_main2}:}
\par The result of Lemma \ref{HankelH0} and Lemma \ref{HankelH1} can be directly used to derive  Theorems \ref{volRes2_main} and \ref{volRes2_main2}. Actually, the matrices $\bm H_1$ and $\bm H_2$ will be replaced by the Hankel matrices $\bm X_1^{[\Delta d\cdot T_s]}$ and $\bm X_2$, then the $\bm H_{1,l_1}$ and $\bm H_{2,l_2}$ are just $\bm S_{1,l_1}^{[\Delta d \cdot T_s]}$ and $\bm S_{2,l_2}$; 
Therefore the expressions (\ref{Hankelcoh01}-\ref{Hankelcoh00}) in Lemma \ref{HankelH0} and (\ref{Hankelcoh10}) in Lemma \ref{HankelH1} are actually related with the auto-correlation function of $s(k\cdot T_s)$, namely, they can be translated into:
\begin{footnotesize}
	\begin{eqnarray}
	\frac{|\langle\bm s_{1,l_1}^{(k)[\Delta \tau]}, \bm s_{2,l_2}^{(k')} \rangle|}{\|\bm s_{1,l_1}^{(k)[\Delta \tau]}\|_2\|\bm s_{2,l_2}^{(k')}\|_2}  &=& \frac{|R_s(\tau_{1,l_1}+\Delta \tau-
		\tau_{2,l_2}+(k-k')T_s)|}{|R_s(0)|}\nonumber\\
	\frac{|\langle \bm s_{1,l_1}^{(k)[\Delta \tau]}, \bm s_{1,{l_1}'}^{(k')[\Delta \tau]} \rangle |}{\|\bm s_{1,l_1}^{(k)[\Delta \tau]}\|_2\|\bm s_{1,{l_1}'}^{(k')[\Delta \tau]}\|_2} &=& \frac{|R_s(\tau_{1,l_1}-\tau_{1,{l_1}'}+(k-k')T_s)|}{|R_s(0)|} \nonumber\\
	\frac{|\langle \bm s_{2,l_2}^{(k)}, \bm s_{2,{l_2}'}^{(k')} \rangle |}{\|\bm s_{2,l_2}^{(k)}\|_2\|\bm s_{2,{l_2}'}^{(k')}\|_2} &=& \frac{|R_s(\tau_{2,l_2}-\tau_{2,{l_2}'}+(k-k')T_s)|}{|R_s(0)|}\nonumber
	\end{eqnarray}
\end{footnotesize}
\par The sufficient condition in (\ref{Hankelmucond0}) and (\ref{Hankelmucond1}) are obtained through similar techniques in the proofs of Theorem \ref{volRes1_main3} and \ref{volRes1_main2}.


\ifCLASSOPTIONcaptionsoff
  \newpage
\fi

\end{document}